\documentclass[a4paper,fleqn,usenatbib]{mnras}
\usepackage{newtxtext,newtxmath}
\usepackage[T1]{fontenc}
\usepackage{ae,aecompl}
\usepackage{booktabs}
\usepackage[labelfont=bf, font=footnotesize]{caption}
\usepackage{multirow, color, colortbl} % Tables with multi rows & coloured rows
\usepackage{graphicx} % Including figure files
\usepackage{subfig} % Subfigures
\usepackage{mathrsfs,bigints} % Advanced maths commands, Extra maths symbols, `Script' maths letters
\usepackage{bm} % Bold italic maths letters
\usepackage{pdflscape} % Landscape pages
\usepackage{version} % comment multiple lines
\newcommand{\rstar}{r_*}
\newcommand{\rstari}{r_{*i}}
\newcommand{\rh}{r_{\rm h}}
\newcommand{\rt}{r_{\rm t}}
\newcommand{\Reff}{R_{\rm e}}
\newcommand{\Rd}{R_{\rm d}}

\newcommand{\phistar}{\phi_*}
\newcommand{\phij}{\phi_j}
\newcommand{\tphij}{\tilde\phi_j}
\newcommand{\phistari}{\phi_{*i}}
\newcommand{\tphistari}{\tilde\phi_{*i}}

\newcommand{\nustar}{\nu_*}

\newcommand{\vphi}{\upsilon_{\varphi}}
\newcommand{\sigphi}{\sigma_{\varphi}}
\newcommand{\sigphii}{\sigma_{\varphi i}}
\newcommand{\sigmi}{\sigma_i}
\newcommand{\sigmij}{\sigma_{ij}}

\newcommand{\ke}{k_{\rm e}}
\newcommand{\Sigmastar}{\Sigma_*}
\newcommand{\vlos}{\upsilon_{\rm los}}
\newcommand{\siglos}{\sigma_{\rm los}}
\newcommand{\sigP}{\sigma_{\rm P}}
\newcommand{\VP}{V_{\rm P}}
\newcommand{\Vrms}{V_{\rm rms}}

\newcommand{\vlosL}{\upsilon_{\mathrm{los}\mathscr{L}}}
\newcommand{\siglosL}{\sigma_{\mathrm{los}\mathscr{L}}}

\newcommand{\Di}{\Delta_i}
\newcommand{\tDi}{\tilde\Delta_i}
\newcommand{\Dij}{\Delta_{ij}}
\newcommand{\tDij}{\tilde\Delta_{ij}}

\newcommand{\tsigmi}{\tilde\sigma_{i}}
\newcommand{\tsigmij}{\tilde\sigma_{ij}}

\newcommand{\tilda}{\tilde a}
\newcommand{\tildb}{\tilde b}
\newcommand{\tildz}{\tilde z}
\newcommand{\tildR}{\tilde R}

\newcommand{\Wi}{W_{i}}
\newcommand{\Wij}{W_{ij}}

\newcommand{\tWij}{\tilde W_{ij}}

\newcommand{\Jzer}{{\rm J}_0}

\newcommand{\RM}{\mathcal{R}}
\newcommand{\RMi}{\RM_i}
\newcommand{\RMj}{\RM_j}
\newcommand{\RMh}{\RM_{\rm h}}
\newcommand{\RMbh}{\RM_{\rm BH}}
\newcommand{\RMo}{\RM_1}
\newcommand{\RMt}{\RM_2}
\newcommand{\RMax}{\RM_{\rm M}}
\newcommand{\RMd}{\RM_{\rm d}}

\newcommand{\Mh}{M_{\rm h}}
\newcommand{\MBH}{M_{\rm BH}}
\newcommand{\Mstar}{M_*}
\newcommand{\Mstari}{M_{*i}}
\newcommand{\Msone}{M_{*1}}
\newcommand{\Mstwo}{M_{*2}}

\newcommand{\rhostar}{\rho_*}
\newcommand{\trhostar}{\tilde\rho_*}
\newcommand{\rhoi}{\rho_{*i}}
\newcommand{\trhoi}{\tilde\rho_{*i}}
\newcommand{\rhof}{\rho_{*1}}
\newcommand{\trhof}{\tilde\rho_{*1}}
\newcommand{\rhos}{\rho_{*2}}
\newcommand{\trhos}{\tilde\rho_{*2}}
\newcommand{\rhoh}{\rho_{\rm h}}
\newcommand{\trhoh}{\tilde\rho_{\rm h}}
\newcommand{\rhon}{\rho_{\rm n}}

\newcommand{\phin}{\phi_{\rm n}}
\newcommand{\phih}{\phi_{\rm h}}
\newcommand{\tphih}{\tilde\phi_{\rm h}}
\newcommand{\phiBH}{\phi_{\rm BH}}
\newcommand{\sumi}{\sum_i}
\newcommand{\sumj}{\sum_j}

\newcommand{\nui}{\nu_{*i}}
\newcommand{\rmh}{\mathrm{h}}
\newcommand{\rmBH}{\mathrm{BH}}

\newcommand{\barvphi}{\overline{\mathrm{v}_\varphi^2}}

\newcommand{\vphii}{\upsilon_{\varphi i}}

\newcommand{\bfn}{\mathbf{n}}
\newcommand{\bfe}{\mathbf{e}}
\newcommand{\evphi}{\bfe_\varphi}
\newcommand{\dl}{\,\mathrm{d}l}
\newcommand{\dt}{\,\mathrm{d}t}
\newcommand{\dlambda}{\,\mathrm{d}\lambda}

\newcommand{\rmlos}{\mathrm{los}}
\newcommand{\dR}{\,\mathrm{d}R}
\newcommand{\dz}{\,\mathrm{d}z}

\newcommand{\mtoli}{\Upsilon_{*i}}
\newcommand{\mtol}{\Upsilon_{*}}
\newcommand{\MLo}{\Upsilon_{*1}}
\newcommand{\MLt}{\Upsilon_{*2}}

\newcommand{\Sigmai}{\Sigma_{*i}}
\newcommand{\rme}{\mathrm{e}}

\defcitealias{b&t08}{BT08}
\defcitealias{cmpz21}{CMPZ21}
\defcitealias{ciot_gan21}{Ciotti}

\title[Jeans modeling with multiple stellar populations]{Jeans modeling of axisymmetric galaxies with multiple stellar populations}

\author[C. Caravita, L. Ciotti and S. Pellegrini]{
	Caterina Caravita, $^{1,2}$ \thanks{E-mail: caterina.caravita2@unibo.it}
	Luca Ciotti $^{1}$
	and Silvia Pellegrini $^{1,2}$
	\\
	$^{1}$ Department of Physics and Astronomy, University of Bologna, via P. Gobetti 93/2, 40129 Bologna, Italy \\
    $^{2}$ INAF-OAS of Bologna, via P. Gobetti 93/3, 40129 Bologna, Italy}

\pubyear{2019}

\begin{document}
\label{firstpage}
\pagerange{\pageref{firstpage}--\pageref{lastpage}}
\maketitle

\begin{abstract}

We present the theoretical framework to efficiently solve the
Jeans equations for multi-component axisymmetric stellar systems,
focusing on the scaling of all quantities entering them. The
models may include an arbitrary number of stellar distributions, a
dark matter halo, and a central supermassive black hole; each
stellar distribution is implicitly described by a two- or three-integral
distribution function, and the stellar components can have different
structural (density profile, flattening, mass, scale-length),
dynamical (rotation, velocity dispersion anisotropy), and population
(age, metallicity, initial mass function, mass-to-light ratio)
properties. In order to determine the ordered rotational velocity
and the azimuthal velocity dispersion fields of each component, we
introduce a decomposition that can be used when the commonly adopted
Satoh decomposition cannot be applied. The scheme developed is
particularly suitable for a numerical implementation; we describe its
realisation within our code JASMINE2, optimised to maximally
exploit the scalings allowed by the Poisson and the Jeans equations,
also in the post-processing procedures. As applications, we illustrate the building of three
multi-component galaxy models with two distinct stellar populations, a
central black hole, and a dark matter halo; we also study the
  solution of the Jeans equations for an exponential thick disc, and
  for its multi-component representation as the superposition of three
  Miyamoto-Nagai discs. A useful general formula for the numerical
  evaluation of the gravitational potential of factorised thick discs
  is finally given.
\end{abstract}


\section{Introduction} \label{sec:intro}

Axisymmetric galaxy models often represent an acceptable description
of real galaxies, beyond the zeroth-order approximation of spherical
symmetry. Analytical models of one and multi-component axisymmetric
galaxies are available (e.g., see \citealt{b&t08}, hereafter BT08, and
references therein; see also \citealt{cmpz21}, hereafter CMPZ21), but
these models, while important to highlight fundamental properties of
the dynamics of axisymmetric systems, and to guide the construction of
realistic galaxy models to be carried out numerically, suffer from the
restrictions imposed by the request of analytical tractability. From
this point of view, analytical and numerical modeling should be seen
as complementary approaches, each of them with their own merits and
limitations.

On the numerical side, the most common models are based on the
solution of the Jeans equations (e.g. \citetalias{b&t08}; \citealt{ciotti21}). This approach allows to model axisymmetric
stellar systems, in the simplest assumption of a 2-integral
phase-space distribution function \citep[DF, e.g.][]{pos13}, or a
3-integral DF \citep[e.g.][]{capp08}, starting from the assignment of
the density components. As well known, the proper description of a
stellar system should start from the assignment of the phase-space DF
of each separate mass component, and the solution of the associated
Poisson equation (\citetalias{b&t08}; \citealt{bertin14,ciotti21});
however, in several applications, the Jeans approach is highly
preferred, for its direct control on the density distributions (even
if it leaves open fundamental issues such as the phase-space
consistency).

In this paper we present a procedure especially designed to build
(and project) multi-component systems.  For an arbitrary number of
mass components, we start with maximally exploiting the scalings
allowed by the Poisson and the Jeans equations, and then we show how to combine the solutions for all components to obtain a particular
multi-component model. When implemented numerically, this scheme
allows for a fast and flexible building of realistic models. To
provide an example, we describe how the various steps of the
procedure were inserted in our code JASMINE \citep[\textit{Jeans
AxiSymmetric Models of galaxies IN Equilibrium},][]{pos13} for the
axisymmetric modeling of galaxies based on the Jeans equations; the resulting much extended code version was named JASMINE2.  This
new code allows for the choice, in input, of different stellar
components and Dark Matter (DM) components (from a continuously
updated library), and a central Black Hole (BH). It computes
numerically the gravitational potential of each density component by
using the well-known formula based on complete elliptic integrals of
the first kind. This
numerical method is highly accurate, but it easily becomes quite time
expensive, depending on the grid resolution and on the number of
density components; if one wants to explore the parameter space
(that can be very large, especially for multi-component models),
the possibility of a scheme to allow for a full scaling of the
Poisson and the Jeans equations is crucial. The building of a
model is then organised in two distinct parts: in the first one, whose numerical realisation we call \textit{Potential and
Jeans Solver}, one computes the potential and then solves the Jeans
equations for each scaled stellar density component; this
produces a set of solutions that represents a ''progenitor'' of a
family of models. In the second part, that in our numerical
realisation is seen as a \textit{Post-Processing} (PP) phase,
the mass and luminosity weights are assigned, and the kinematical
decompositions imposed; with these, the scaled solutions of the
progenitor are finally combined, and the resulting kinematical
fields projected.

This procedure allows to drastically reduce the computational time
needed for the construction of a multi-component model: with a single
run of the Potential and Jeans Solver, one can build a {\it family} of
galaxy models, all characterised by the same set of scaled density
components; each specific model in the family is defined by choosing
suitable weights and kinematical decompositions in PP. In this way the
exploration of the parameter space is extremely fast and
complete. Summarising, each stellar density component in a
multi-component model is characterised by different structural
(density profile, flattening, total mass, scale-length), dynamical
(rotational support, velocity dispersion anisotropy), and stellar
population (age, metallicity, initial mass function, mass-to-light
ratio) properties. The addition of a central BH and a DM halo is
immediate.

The paper is organised as follows. Section \ref{sec:multi-comp}
presents the analytical framework of the procedure, together
with a new velocity decomposition for the azimuthal velocity field, to
be used when the commonly adopted \citet{satoh80} $k$-decomposition
cannot be applied (a not uncommon case in multi-component systems).
In Section \ref{sec:scaling} we detail how the scaling is carried out.  In Section \ref{sec:app} some illustrative galaxy models are built, and a few tests are mentioned; we also present an application to the case of the exponential disc and its decomposition as sum of Miyamoto-Nagai discs. In Section \ref{sec:concl} the main conclusions are summarised. Finally, Appendix \ref{app:pos_cond} and \ref{app:expdisc} contain some relevant analytical details.

\section{Multi-component galaxy models} \label{sec:multi-comp}
In this section we introduce the general notation used, and we illustrate the main theoretical foundations on
which our modeling procedure is based; in particular, we focus on the multi-component Jeans equations, on a generalisation of the Satoh $k$-decomposition for azimuthal motions, and on the projections on the
plane of the sky.  In the following Section \ref{sec:scaling}, we will describe the scaling procedure, and in particular how the scaled
solutions of the Jeans equations are obtained (with the Potential and
Jeans Solver), and then combined by adopting suitable weights (in the
PP phase).

\subsection{Structure of the galaxy models}
\label{sec:structure}

We adopt cylindrical coordinates $(R,\varphi,z)$, with the
symmetry axis of the models aligned with the $z$-axis.  In full
generality, we consider models composed of $N$ different stellar
density distributions $\rhoi (R,z)$, of total mass $\Mstari$, so that
the total stellar density $\rhostar$ and the total stellar mass
$\Mstar$ of the system are given respectively by
\begin{equation}
  \rhostar(R,z) = \sumi \rhoi, \quad \Mstar = \sumi \Mstari,\quad i=1,\ldots,N.
\label{eq:rho_sum}
\end{equation}
From now on, sums over $i$ indicate sums over the $N$
stellar components.  We assume that each $\rhoi$ is made of a simple
stellar population \citep[see e.g.][]{renz-buz86,mar05}, i.e. by
stars of the same age, chemical composition, initial mass function,
and in particular the same mass-to-light ratio $\mtoli$.  Therefore,
the total stellar distribution $\rhostar$ can be considered a composite
stellar population; the luminosity density and the total
luminosity of each stellar component can be written respectively as
\begin{equation}
 \nui(R,z) = {\rhoi \over \mtoli}, \quad L_i = {\Mstari \over \mtoli},
  \label{eq:nui_Li}
\end{equation}
so that
\begin{equation}
 \nustar(R,z) = \sumi\nui, \quad L = \sumi L_i.
\end{equation}
The local and average stellar mass-to-light ratios of the galaxy are given by
\begin{equation} 
  \mtol(R,z)\equiv {\rhostar\over\nustar}={\sumi
    \rhoi\over\sumi\rhoi/\mtoli},\quad
  <\mtol> \equiv {\Mstar\over L} = {\sumi\Mstari\over\sumi\Mstari/\mtoli},
  \label{eq:mtol}
\end{equation}
where it is apparent how in general the local stellar mass-to-light ratio 
in a multi-component model depends on position.

From equation \eqref{eq:rho_sum} the gravitational potential
associated with the total stellar density is
\begin{equation} 
  \phistar(R,z) = \sumi \phistari,
  \label{eq:phi}
\end{equation}
where $\phistari (R,z)$ is the potential originated by the density
component $\rhoi$.  The presence of a central BH, of mass $\MBH$,
produces the potential
\begin{equation}
\phiBH(r) = -{G\MBH\over r}, \quad r=\sqrt{R^2+z^2},
   \label{eq:phi_bh}
\end{equation}
and an axisymmetric DM halo, of density $\rhoh(R,z)$ and total mass $\Mh$ 
(when finite), produces the potential $\phih(R,z)$. Therefore, in general,
the total gravitational potential of the model is
\begin{equation} 
  \Phi(R,z) = \phistar + \phih + \phiBH = \sumj \phij,\quad j=1,\ldots,N+2.
  \label{eq:phi_sum}
\end{equation}
From now on, sums over $j$ indicate sums over \textit{all} the $N+2$
galaxy components, i.e. the $N$ stellar components, the central BH,
and the DM halo. In principle, also the DM distribution can be made of
different components, with a trivial generalisation of the current
discussion, that is not necessary for the goal of this paper. Our scheme
fully exploits the linearity of the Jeans equations with respect to
the stellar density (Section \ref{sec:JEs}) and to the gravitational potential (Section \ref{sec:scaling}).

\subsection{ The Jeans equations}
\label{sec:JEs}

The procedure, in its basic version, assumes that each stellar component is
implicitly described by a 2-integral phase-space DF $f_i(E,J_z)$ (in
general different for each component), where $E$ and $J_z$ are
respectively the energy and the axial component of the angular
momentum of each star (per unit mass) in the {\it total} potential
$\Phi$. Therefore, the DF of the total stellar distribution is the
2-integral function
\begin{equation}
  f=\sumi f_i.
  \label{eq:DF_sum}
\end{equation}
As usual, we indicate with $({\rm v}_R, {\rm v}_\varphi, {\rm v}_z)$
the velocity components in the phase-space, and with a bar over a
quantity the operation of average over the velocity-space. By
construction, for each stellar component
$\overline{{\rm v}_{R}}_i=\overline{{\rm v}_{z}}_i=0$, the only
non-zero ordered velocity can
occur in the azimuthal direction $\vphii \equiv \overline{{\rm v}_\varphi}_i$, and finally for the velocity
dispersion tensor $\sigma_{Ri}=\sigma_{zi}\equiv \sigmi$. Of course,
from equation \eqref{eq:DF_sum} similar relations hold for the
kinematical fields of the total $\rhostar$.

The Jeans equations for each stellar component are obtained as
velocity averages of the Collisionless Boltzmann Equation over the
corresponding $f_i$ (e.g. \citetalias{b&t08}), so that
\begin{equation}
 \begin{cases}
  \displaystyle{{\partial\rhoi\sigmi^2\over\partial z} = -\rhoi {\partial \Phi\over\partial z}}, \cr\cr 
  \displaystyle{{\partial\rhoi\sigmi^2\over\partial R} = \rhoi {\Di\over R} -\rhoi {\partial \Phi\over\partial R}}, 
\end{cases}
 \label{eq:JEs}
\end{equation}
where $\Phi$ is the total potential in equation \eqref{eq:phi_sum},
and
\begin{equation}
 \Di\equiv \barvphi_i-\sigmi^2,\quad
 \sigphii^2 =\barvphi_i-\vphii^2=\Di + \sigmi^2-\vphii^2;
   \label{eq:pzi_Deltai_pphii}
 \end{equation}
 therefore, in the isotropic case, $\Di = \vphii^2$. Imposing the
 natural boundary condition $\rhoi\sigmi^2\to 0$ for $z\to\infty$, the
 solution of equations \eqref{eq:JEs} is
\begin{equation} 
\rhoi\sigmi^2 = \int_z^\infty \rhoi {\partial \Phi\over\partial z'} dz',\quad
 \rhoi\Di = R \bigg({\partial\rhoi\sigmi^2\over\partial R}+\rhoi {\partial \Phi\over\partial R}\bigg).
 \label{eq:JEsi_sol}
\end{equation}
We notice that $\Di$ can be also recast as a commutator-like integral
(e.g., see equation 35 in \citetalias{cmpz21}), with some advantage
for analytical and numerical investigations; however we found by
several numerical tests that $\Di$ can also be accurately computed by
(centred) numerical differentiation as in equation
\eqref{eq:JEsi_sol}, and so in our code JASMINE2 we maintained this more direct way of evaluation.

A central point of the procedure is the sum rule in 
the
phase-space imposed by the identity \eqref{eq:DF_sum}.  In fact, with
equation \eqref{eq:DF_sum}, we are assuming that the $N$ stellar
components $\rhoi$ are physically distinct, each of them described by
its own $f_i$, and so necessarily the $N$ pairs of equations
\eqref{eq:JEs} are the moment equations of each $f_i$ in the total
potential $\Phi$.  As usual, if $F(\mathbf{x},\mathbf{v})$ is
a generic dynamical property defined over the phase-space, then
\begin{equation} 
  \overline{F}_i = {\int F f_i d^3{\bf v}\over\rhoi},\quad
    \overline{F} = {\sumi \rhoi \overline{F}_i\over\rhostar},\quad 
  \overline{F}_\mathscr{L} = {\sumi \nui \overline{F}_i\over\nustar},
  \label{eq:lw_general}
\end{equation}
where the properties $\overline{F}$ and $\overline{F}_\mathscr{L}$ of
the galaxy can be interpreted as the
\textit{mass-weighted} and the \textit{luminosity-weighted} averages
of the $\overline{F}_i$, respectively. Notice that we are not reconstructing here the phase-space DFs of the models; we just determine the general rules of combination of the velocity moments in multi-component systems. The previous considerations show how
to combine the solution for the single $\rhoi$ to obtain the
dynamical fields associated with the total $\rhostar$.  Clearly, the
Jeans equations for $\rhostar$ in equation \eqref{eq:rho_sum} are
obtained as the sum of equations \eqref{eq:JEs} over the $N$
components, and their solution can be written as
\begin{equation}
 \begin{cases}
\displaystyle{\sigma^2 = {\sumi \rhoi\sigmi^2\over\rhostar}, \quad \Delta = {\sumi \rhoi\Di\over\rhostar},}\\
\displaystyle{
\barvphi = {\sumi \rhoi \barvphi_i\over\rhostar},\quad \vphi = {\sumi
    \rhoi \vphii\over\rhostar}},
\end{cases}
\label{eq:pz_Delta_rhvphi_sum}
\end{equation}
where the previous identities are of straightforward proof from
equations \eqref{eq:lw_general} and \eqref{eq:pzi_Deltai_pphii}.  Note that $\sigphi^2$ is \textit{not}
given by the simple sum of the $\sigphii^2 $ of the single
components, as $\sigma^2$ in equation \eqref{eq:pz_Delta_rhvphi_sum}
above, because from equations \eqref{eq:pz_Delta_rhvphi_sum} and
\eqref{eq:pzi_Deltai_pphii} one has:
\begin{equation} 
\sigphi^2 =\barvphi-\vphi^2= \Delta +\sigma^2- \vphi^2=
{\sumi \rhoi (\sigphii^2 +\vphii^2)\over\rhostar} -\vphi^2. 
\label{eq:pphi_sum}
\end{equation}
Similarly, from equation \eqref{eq:lw_general}, we derive all the
corresponding luminosity-weighted quantities; we do not give here
their expressions, since they are just obtained by using as weights
the luminosity densities $\nui $ of the components instead of the mass
densities $\rhoi$, in equations \eqref{eq:pz_Delta_rhvphi_sum} and
\eqref{eq:pphi_sum}.

Finally, the rotation curve in the equatorial plane is given in terms of the circular velocity $\upsilon_{{\rm c}j}$ of each mass component as
\begin{equation} 
\upsilon_{\rm c}^2=\sumj \upsilon_{{\rm c}j}^2.
\label{eq:vcsum}
\end{equation}

\subsection{Azimuthal velocity decomposition}
\label{sec:satoh}

As well known, equations \eqref{eq:JEs} are degenerate in the
azimuthal direction, i.e. they only provide
$\barvphi_i=\sigphii^2 + \vphii^2$. The most common phenomenological
approach to break this degeneracy is the \citet{satoh80}
$k$-decomposition. If $\Di\geq 0$ over the whole space, then one
assumes
\begin{equation} 
 \vphii = k_i \sqrt{\Di},\qquad \sigphii^2 =\sigma_i^2+\big(1-k_i^2\big)\Di,
 \label{eq:satoh_i}
\end{equation}
with negative values of $k_i$ describing clockwise rotation. The
special case $k_i^2=1$ corresponds to the isotropic rotator
($\sigphii=\sigma_i$), with flattening totally supported by rotation;
while, if $k_i=0$, there is no net rotation ($\vphii=0$), and the
flattening is totally supported by tangential velocity
anisotropy. More general velocity decompositions can be obtained by
assuming a position-dependent parameter $k_i(R,z)$, also allowing for
values greater than unity, up to a position-dependent maximum
determined by the request $\sigphii=0$ \citep[e.g.][]{satoh80,
  ciot-pel96, negri14}. In principle each stellar component of a
multi-component model is characterised by a different $k_i$, so that
from equations \eqref{eq:satoh_i} and \eqref{eq:pz_Delta_rhvphi_sum}
the total $\rhostar$ will have an \textit{effective} Satoh parameter
$\ke$ given by
\begin{equation}
\ke\equiv {\vphi\over\sqrt{\Delta}}={\sumi
  k_i\rhoi\sqrt{\Di}\over\rhostar\sqrt{\Delta}},\quad
\sigphi^2 = \sigma^2+\big(1-\ke^2\big)\Delta,
\label{eq:satoh_e}
\end{equation}
where $\ke$ in general depends on position, even if the $k_i$ do not.

Clearly, in case of $\Di <0$ for some $\rhoi$, the Satoh decomposition
in equation \eqref{eq:satoh_i} cannot be applied. The case of a
negative $\Di$ over some regions of space (or everywhere) is not
frequently encountered in applications, but it is not impossible; for
example, it necessarily occurs for density components in
multi-component systems with spherically symmetric total density, or
in density distributions elongated along the symmetry axis \citep[see
e.g. Chapter 13 in][]{ciotti21}. Indeed, $\Delta=0$ everywhere for a
spherical system supported by a 2-integral DF, and thus, from equation
\eqref{eq:pz_Delta_rhvphi_sum}, at least one $\Di$ must be negative
(excluding the trivial case of all the subcomponents spherically
symmetric, so that $\Di=0$). Notice that $\Di <0$ is not necessarily a
manifestation of an inconsistent DF ($f_i<0$), while if
$\barvphi_i = \Di+\sigma_i^2<0$ \textit{certainly} the whole model must be discarded as unphysical, even if the solution for the \textit{total} stellar distribution are well-behaved. Therefore, if for some component $\Di<0$, then in case of positivity of the sum $\Di+\sigma_i^2\geq 0$, the Satoh decomposition is generalised to:
\begin{equation} 
  \vphii = k_i \sqrt{\Di+\sigma_i^2},\quad \sigphii^2 =
  \big(1-k_i^2\big)\big(\Di+\sigma_i^2\big),\quad k_i^2\leq 1,
  \label{eq:g-satoh_i}
\end{equation}
where again $k_i$ can depend on position. We refer to this alternative
decomposition as to the {\it generalised} $k$-decomposition.  The case
$k_i=0$ implies no net rotation ($\vphii=0$), while now $k_i^2=1$
corresponds to $\sigphii=0$; notice that no isotropic rotators can be
realised from equation \eqref{eq:g-satoh_i} when $\Di<0$, because
isotropy ($\sigphii=\sigma_i$) would correspond to $k_i^2 <
0$. Moreover, while with the Satoh decomposition a spherical system
cannot rotate and is isotropic independently of the value of $k_i$,
with the generalised $k$-decomposition one can model rotating (and
anisotropic) spherical systems. An application of this last case can
be found, for example, in exploratory numerical simulations of rotating gas flows in
galaxies of \citet{Yoon19}. Notice that the generalised decomposition
applied to systems with $\Di \gg \sigma_i^2$ (as for instance the case
of highly flattened discs) reduces to the standard Satoh formula. A
more interesting (and delicate) case, requiring particular care in the
choice of the parameter $k_i$, is represented by systems with
$|\Di|\ll \sigma_i^2$, when we have $\vphii \sim k_i \sigma_i$, and
$\sigphii^2 \sim (1-k_i^2) \sigma_i^2$. This means that, in order to
avoid substantial rotation, for example in almost spherical systems
(oblate or prolate), $k_i$ must be kept small.

We finally remark that, for a given multi-component system, it is also
possible to assume a Satoh decomposition for some components, and the
generalised decomposition for the others; in analogy with equation
\eqref{eq:satoh_e} it is possible to define a total \textit{effective}
decomposition parameter $\ke$ as
\begin{equation}
 \ke\equiv {\vphi\over \sqrt{\Delta+\sigma^2}},\quad
  \sigphi^2 = \big(1-\ke^2\big)\big(\Delta+\sigma^2\big)\,.
 \label{eq:g-satoh_e}
\end{equation}

\subsection{Projections}
\label{sec:proj}

We recast here the projection formulae presented in \citet{pos13} for
the case of a multi-component system, focusing in particular on how the
solutions for the components must be summed to obtain the projected
fields of the total stellar distribution.  We indicate with $<,>$ the
scalar product, with $\bfn$ the line-of-sight direction (hereafter
los) directed from the observer to the galaxy\footnote{At variance
  with the convention adopted in \citetalias{cmpz21}, where $\bfn$ points 
from the
  galaxy to the observer.}, and with $l$ the integration path along
the los. For the ease of notation in this Section we drop the subscript $i$, so that
all the following formulae must be intended to hold separately for
each stellar component $\rhoi$ (and of course also for the total
$\rhostar$). We will resume the use of the subscript $i$ at the end of the
Section, when we give the expressions for the projected fields of
$\rhostar$ as functions of the projected fields of the components. The projection of a stellar density,
and of the ordered velocity $\bm{\upsilon}=\vphi\,\evphi$, are
\begin{equation}
  \Sigmastar = \int_{-\infty}^{\infty}\rhostar \dl, \quad
  \Sigmastar\vlos = \int_{-\infty}^{\infty}\rhostar\vphi <\evphi,\bfn> \dl,
\label{eq:sigma_vlos}
\end{equation}
where $\evphi = (-\sin\varphi,\cos\varphi,0)$ is the unitary vector in
the tangential direction. From the adopted orientation of $\bfn$, a
positive/negative $\vlos$ indicates a motion receding from/approaching
to the observer, respectively. The los velocity dispersion can be
written as
\begin{equation} 
 \siglos^2 = \sigP^2+\VP^2-\vlos^2 = \Vrms^2-\vlos^2  
\label{eq:sigmalos}
\end{equation}
(e.g. \citealt{ciot-pel96,pos13}, \citetalias{cmpz21}), where following \citet{capp08} we also define $\Vrms^2 \equiv \sigP^2+\VP^2$, and
\begin{equation}
 \Sigmastar\sigP^2 = \int_{-\infty}^{\infty}\rhostar <\mathbf{\sigma}^2\bfn, \bfn>\dl,
\label{eq:sigp}
\end{equation}
\begin{equation}
 \Sigmastar\VP^2 = \int_{-\infty}^{\infty}\rhostar \vphi^2 <\evphi,\bfn>^2\dl,
\label{eq:vp}
\end{equation}
where in equation \eqref{eq:sigp} $\mathbf{\sigma}$ is the $3\times 3$
velocity dispersion tensor. The fields $\Vrms$ and $\vlos$ in general
depend on the specific direction $\bfn$, and $\vlos$, $\sigP$ and
$\VP$ on the specific velocity decomposition adopted, but $\Vrms$ is
independent of the velocity decomposition. The previous identities are
fully general and hold for a generic inclination of the los with
respect to the galaxy. For our axisymmetric models, it is
assumed without loss of generality that the los is parallel to the
$x-z$ plane, and the projection plane rotates around the $y$ axis.

In particular, in the face-on projection (hereafter FO), the los is parallel to the
$z$ axis with $\bfn=-\bfe_z$, the projection plane is the $x-y$ plane,
and
\begin{equation}
  \Sigmastar = 2\int_0^{\infty}\rhostar \dz, \quad \Sigmastar\siglos^2 = 2\int_0^{\infty}\rhostar\sigma^2 \dz, 
\end{equation}
because $\vlos=\VP=0$, and so $\siglos=\sigP$. In the edge-on projection (hereafter EO), the los is aligned with the
$x$ axis with $\bfn=-\bfe_x$, the projection plane coincides with the
$y-z$ plane, and $(\cos\varphi, \sin\varphi) = (x/R, y/R)$ where
$R=\sqrt{x^2+y^2}$. Then, from equation \eqref{eq:sigma_vlos},
\begin{equation} 
\Sigmastar = 2 \int_y^{\infty}{\rhostar R\over\sqrt{R^2-y^2}}\dR, \quad
\Sigmastar\vlos = 2y
\int_y^{\infty}{\rhostar\vphi\over\sqrt{R^2-y^2}}\dR .
\label{eq:sigmavlos}
\end{equation}
Moreover, with some algebra, from equations \eqref{eq:sigp} and
\eqref{eq:vp} we have
\begin{equation}
\Sigmastar\sigP^2 = 2 \int_y^{\infty}{( R^2-y^2)\sigma^2 + y^2\sigma_{\varphi}^2\over R\sqrt{R^2-y^2}}\rhostar \dR ,
\end{equation}
\begin{equation}
 \Sigmastar\VP^2 = 2y^2 \int_y^{\infty} {\rhostar\vphi^2\over R\sqrt{R^2-y^2}}\dR ,
\end{equation}
so that, from equation \eqref{eq:pzi_Deltai_pphii}, equation
\eqref{eq:sigmalos} can be recast in compact form as
\begin{equation}
\Sigmastar\siglos^2 = 2\int_y^{\infty} {R^2\sigma^2 + y^2\Delta\over R\sqrt{R^2-y^2}}\rhostar \dR - \Sigmastar\vlos^2,
\label{eq:siglos} 
\end{equation}
where the independence of $\Vrms$ from the specific azimuthal velocity
decomposition is apparent.

The projection formulae for a multi-component stellar system can now be
easily obtained, for a generic los, just by considering how the intrinsic 
quantities
add. From equations \eqref{eq:rho_sum} and \eqref{eq:pz_Delta_rhvphi_sum}, and
from equations~\eqref{eq:sigma_vlos}--\eqref{eq:vp}, it is immediate
to see that
\begin{equation} 
  \Sigmastar = \sumi \Sigmai ,\quad
  \vlos = {\sumi \Sigmai\,\upsilon_{\mathrm{los}i}\over\Sigmastar},\quad
   \Vrms^2 = {\sumi\Sigmai V_{\mathrm{rms}i}^2\over\Sigmastar},
  \label{eq:sigmavlosi_sumi}
\end{equation}
and $\siglos^2$ is given again by equation \eqref{eq:sigmalos}.

No difficulty is encountered in the construction of the
luminosity-weighted fields analogous to
equations~\eqref{eq:sigmavlosi_sumi}, by using the surface brightness
distributions $I_{*i}=\Sigmai/\mtoli$, and $I_*=\sumi I_{*i}$, so the
{\it projected} stellar mass-to-light ratio
$\Upsilon_{*\rmlos} \equiv \Sigma_*/I_*$, defined in analogy with the
local $\Upsilon_*$ in equation \eqref{eq:mtol}.

Summarising, we now have the framework needed to determine the
solution of the Jeans equations once the solutions for the single
components in the total potential are known. In the following Section
we detail how the solution of each stellar component is obtained,
thanks to the adopted scaling procedure.

\section{Scaling of multi-component models}
\label{sec:scaling}

We show here how, thanks to the full use of the scalings allowed by the Poisson and the Jeans equations, once a set of solutions is obtained for them, one can build an arbitrarily large family of
models, just by combining the scaled solutions in this set with different weights; the scheme thus provides several galaxy models with almost no effort. The
basic idea is elementary. We recognise that equations \eqref{eq:JEs}, at {\it fixed} total potential $\Phi$, are invariant for a mass scaling of the density $\rhoi$, i.e. at fixed $\Phi$ the derived velocity fields would be independent of the value of $\Mstari$. However, as $\Phi$ contains also $\phistari$, equations
\eqref{eq:JEs} obviously are {\it not} invariant to such scaling; nonetheless, the $N+2$ equations for $\rhoi$ in the potentials $\phij$
\begin{equation} 
 \begin{cases}
  \displaystyle{{\partial\rhoi\sigmij^2\over\partial z} = -\rhoi {\partial \phij\over\partial z}}, \cr\cr 
  \displaystyle{{\partial\rhoi\sigmij^2\over\partial R} = \rhoi {\Dij\over R}
    -\rhoi {\partial\phij\over\partial R}},
  \label{eq:JEsij}
 \end{cases} 
\end{equation}
and their solutions
\begin{equation}
 \displaystyle{\rhoi\sigmij^2= \int_z^\infty \rhoi {\partial \phij\over\partial z'}\,\mathrm{d}z',} \quad
\displaystyle{\rhoi\Dij = R\left({\partial\rhoi\sigmij^2\over\partial R}+\rhoi
   {\partial\phij\over\partial R}\right),}
 \label{eq:JEsij_sol}
\end{equation}
with $\rhoi\sigmij^2\to 0$ for $z\to\infty$, do have important scaling
properties that will be exploited in Section
\ref{sec:scaling_scheme}. We note that here and in the following the
double subscript in $\sigmij^2$ does not refer to the tensorial nature
of the velocity dispersion, but just identifies the solution of the
$i$-th stellar component in the $j$-th potential component.

Leaving aside for the moment the scaling properties of
equations \eqref{eq:JEsij} and \eqref{eq:JEsij_sol}, it is obvious that the sums
\begin{equation} 
  \sigmi^2 = \sumj \sigmij^2,\quad \Di = \sumj \Dij, 
  \label{eq:pzi_deltai_sum}
\end{equation}
are the solution of equations \eqref{eq:JEs}, as can be demonstrated,
first by summing over $j$ the $N+2$ equations \eqref{eq:JEsij} and
their solutions \eqref{eq:JEsij_sol}, and comparing the resulting
expressions with equations \eqref{eq:JEs} and \eqref{eq:JEsi_sol}, and
then by proving that the solution of equation \eqref{eq:JEs} is unique
from the imposed boundaries.

An important point is in order here. Despite the apparent similarity
of the decomposition of $\sigmi^2$ and $\Di$ performed in equations
\eqref{eq:JEsij} over the $N+2$ potential components $\phij$, with the
decomposition of $\sigma^2$ and $\Delta$ performed in equations
\eqref{eq:JEs} over the $N$ stellar components $\rhoi$, there is a
fundamental conceptual difference between the two decompositions. In
fact, equations \eqref{eq:JEs} are {\it true} moments of the
Collisionless Boltzmann Equation obeyed by the distribution functions
$f_i$ in the total potential, and so they have a sort of autonomous
physical meaning; equations \eqref{eq:JEsij}, instead, are just a
mathematical decomposition over the different $\phij$ of the Jeans
equations for $\rhoi$. As a consequence, phase-space consistency
arguments apply to the solution of equations \eqref{eq:JEs}, but not
to $\sigma_{ij}^2$ and $\Dij$ separately: as far as the fields
$\sigmi^2$ and $\Di$ are physically acceptable, the model is also
acceptable, independently of the specific properties\footnote{The
  situation is somewhat similar to that faced when decomposing a
  positive density distribution over some prescribed set of functions
  (e.g, spherical harmonics), when the basis functions can present
  regions of negative densities.} of its components $\sigmij^2$ and
$\Dij$.

Finally, we recall the decomposition rule for the Virial Theorem (in its scalar form; the formulae can be easily extended to its tensorial form) of each stellar component:
\begin{equation}  
2K_{*i}=-\Wi=-\sumj\Wij,
\end{equation} 
where $K_{*i}$ is the kinetic energy of the $i$-th stellar component, and
\begin{equation} 
\Wij=-4\pi
G\int_0^{\infty}\int_0^{\infty}\rhoi\left(
  R{\partial\phij\over\partial R} + z{\partial\phij\over\partial
    z}\right)R\,\mathrm{d}R \,\mathrm{d}z.
\end{equation}

\subsection{The scaling scheme}
\label{sec:scaling_scheme}

We describe below how the scaling scheme works in general, with particular reference to its numerical implementation in JASMINE2, and to its logically distinct parts of the Potential and Jeans Solver and of
the PP. We distinguish three groups of model
parameters for the construction of a multi-component model, summarised in Table \ref{tab:fixed_scal_par}.  In the first group there are the {\it physical scales} $\Mstar$ and $\rstar$, i.e. the total stellar mass
and its scale-length. All the density and potential components are made dimensionless by scaling them to the quantities
\begin{equation} 
  \rhon \equiv {\Mstar\over 4\pi \rstar^3}, \quad
  \phin \equiv {G\Mstar\over\rstar}.
  \label{eq:rhon_phin}
\end{equation}
We note that it is convenient to normalise the 2D numerical grid to $\rstar$, with $\tilde R \equiv R/\rstar$, and
$\tilde z \equiv z/\rstar$. A scaled grid guarantees the same resolution, independently of the actual physical 
size of the model, measured by $\rstar$. Incidentally,  JASMINE2
  has a bilogarithmic grid, with a few hundreds
of points in $\tilde R$ and $\tilde z$, ranging from $\approx 10^{-5}$
or less at the origin, up to $\approx 10^2$ or more at the outer edge.
{\it Even though the physical scales are logically introduced first, the values of $\Mstar$ and $\rstar$ (and so of $\rhon$ and $\phin$) are
  fixed in the last step of the model construction, at the end of the PP (see Table \ref{tab:fixed_scal_par})}. In this way, different physical
realisations (in size and total mass) can be obtained for the same multi-component galaxy model.

In the second group of parameters there are the relative mass weights $\RMi\equiv \Mstari/\Mstar$, $\RMh\equiv\Mh/\Mstar$,
$\RMbh\equiv \MBH/\Mstar$ of the different components, the mass-to-light ratios $\mtoli$, and the parameters
$k_i$ appearing in equations \eqref{eq:satoh_i} and
\eqref{eq:g-satoh_i} for the kinematical decomposition of the azimuthal motions. By definition
\begin{equation}
  \sumi\RMi=1,
  \label{eq:sumRi}
\end{equation}
and in full generality we write
\begin{equation} 
  \rhoi = \rhon\RMi\trhoi, \quad 
  \rhoh = \rhon\RMh\trhoh, \quad
   \tilde\rho_{*} = \sumi\RMi\trhoi. 
  \label{eq:rho_tilde}
\end{equation}
where $\trhoi$ and $\trhoh$ are the \textit{scaled} density
distributions, and $\tilde\rho_{*} = \rho_{*}/\rhon$ is the
dimensionless total stellar density. Notice that from equation
\eqref{eq:rhon_phin} the volume integrals of $\trhoi$ over the whole
dimensionless numerical grid evaluate to $4\pi$ by
construction. Similarly,
\begin{equation} 
  \phistari = \phin\RMi\tphistari,\;
  \phih =\phin\RMh\tilde\phih,\;
  \phiBH = \phin\RMbh\tilde\phi_\rmBH,
 \label{eq:phi_tilde}
\end{equation}
and so
\begin{equation}
     \Phi =\phin \sumj\RMj\tphij,\quad 
  \Wi =\Mstar\phin\RMi\sumj\RMj\tWij, 
  \label{eq:W_tilde}
\end{equation}
and
\begin{equation}
  \sigmi^2= \phin\tsigmi^2=\phin \sumj\RMj\tsigmij^2,\quad 
  \Di = \phin\tDi=\phin \sumj\RMj\tDij. 
  \label{eq:kin_tilde}
\end{equation}

Finally, from the assumption of a constant mass-to-light ratio $\mtoli$ for each stellar component,
\begin{equation}
  \nui = \rhon{\RMi\over\mtoli}\trhoi, \quad
  L_i = {\RMi\over\mtoli} \Mstar .
\end{equation}
{\it The values of the weights are chosen in PP (see Table \ref{tab:fixed_scal_par}), because a change in their values, and in the kinematical decompositions, does not require to recompute the potentials and solve again 
  the Jeans equations.} This possibility allows for a fast construction of different models belonging to the same family. 

One family indeed is characterised by the choice of the third group of
parameters, to be performed at the beginning of the model
construction: the {\it structural parameters} of the scaled density
components $\trhoi$ and $\trhoh$, that in full generality we indicate
with the symbols $\xi_i\equiv \rstari/\rstar$ and
$\xi_\rmh\equiv r_\rmh/\rstar$ for the different scale-lengths, and
with $q_i$ and $q_\rmh$ for other parameters that determine the shape
of the scaled densities (for example the flattenings in case of
ellipsoidal density distributions). {\it The values of the structural
  parameters must be assigned in order to run the Potential and  Jeans
  Solver (see Table \ref{tab:fixed_scal_par}), and in general a change
  in some of their values requires a new computation of the potentials
  and of the Jeans solutions.}

%%%%%%%%%%%%%%%%%%%%%%%%%%%%%%%%%%%%%%%%%%%%%%%%%%%%%%%%%%%%
\begin{table}
 \centering 
 \begin{tabular}{ll}
     \multicolumn{2}{c}{Parameters of the scaling scheme} \\
  \toprule\toprule 
   \multicolumn{2}{c}{Potential and Jeans Solver} \\
Structural parameters & \\
   \cmidrule(lr){1-2}
Scaled stellar and DM densities & $\trhoi,\; \tilde\rhoh$ \\
Scale-length ratios & $\displaystyle{\xi_i={\rstari\over\rstar},\, \xi_{\rm h}={r_{\rm h}\over\rstar},\dots}$ \\
Shape parameters & $q_i,\; q_{\rm h},\ldots$ \\
  \midrule\midrule 
  \multicolumn{2}{c}{Post-Processing} \\
Weights & \\
   \cmidrule(lr){1-2}
Mass ratios & $\displaystyle{\RMi = {\Mstari\over\Mstar},\; \RMh={\Mh\over\Mstar},\; \RMbh ={\MBH\over\Mstar}}$ \\
Mass-to-light ratios &  $\displaystyle{\mtoli ={\Mstari\over L_i}}$ \\
   Kinematical decompositions &  $k_i,\;\lambda_i,\;\delta_i$ \\
     \midrule\midrule 
  \multicolumn{2}{c}{Post-Processing} \\
Physical Scales & \\
  \cmidrule(lr){1-2}
Total stellar mass &  $\Mstar$ \\
Total stellar density scale-length & $\rstar$ \\
  \bottomrule\bottomrule 
 \end{tabular}
 \caption{The three main steps involved in the construction of a multi-component model, listed from top to bottom in the order in which they
   are considered in a numerical implementation of the scaling scheme (as described in Section~\ref{sec:scaling_scheme}).}
 \label{tab:fixed_scal_par}
\end{table}
%%%%%%%%%%%%%%%%%%%%%%%%%%%%%%%%%%%%%%%%%%%%%%%%%%%%%%%%%%%%

\subsubsection{The Potential and Jeans Solver}

For a chosen set of values for the structural parameters, the scaled
Jeans equations are obtained from equations \eqref{eq:JEsij} and
equations \eqref{eq:rho_tilde} and \eqref{eq:phi_tilde}. In practice,
for $N$ assigned scaled stellar components $\trhoi$ and a scaled dark
matter halo $\trhoh$, the Potential \& Jeans Solver first computes the
scaled potentials $\tphistari$ and $\tphih$, and then solves the
$N\times (N+2)$ pairs of scaled equations \eqref{eq:JEsij}, one for
each $\trhoi$ in the potential $\tphij$ (including
$\tilde\phi_\rmBH$), over the dimensionless grid
$(\tilde R, \tilde z)$; thus the scaled fields $\tsigmij^2$ and
$\tDij$ are obtained. The possibility to solve equations
\eqref{eq:JEsij} without choosing $\RMi$ and $\RMj$ is due to the fact
that, on one hand, the weights $\RMi$ appear linearly in both sides of
equations \eqref{eq:JEsij}; on the other hand, $\tsigmij^2$ and
$\tDij$ scale linearly with $\RMj$, once the boundary condition is
fixed to zero at infinity.

The details of the numerical implementation of the computation of the
potentials and of the solution of the Jeans equations are described
in \citet{pos13}. Here we recall that the standard choice for the numerical computation of the potential in JASMINE2 is the integral formula
 \begin{equation}
\phi=-4G\int_0^{\infty} R'\,\mathrm{d}R'\int_{-\infty}^{\infty}{\rho(R',z')\,\mathrm{d}z'\over\sqrt{(R+R')^2+\Delta 
    z^2}}{\bf K}\left[
\sqrt{{4RR'\over (R+R')^2+\Delta z^2}}
  \right]
\label{eq:ellpot}
   \end{equation}
where $\Delta z= z -z'$, and ${\bf K}$ is the complete elliptic integral of the first kind (see e.g. \citetalias{b&t08}; \citealt{ciotti21}), evaluated as a 2-dimensional integration over a staggered grid. However, for genuinely ellipsoidal models, the code can the use the faster Chandrasekhar formula (e.g. equation 2.140 in \citetalias{b&t08}; equation 2.21 in \citealt{ciotti21}), and, for disc distributions, the integral formula based on Bessel functions (see Section \ref{sec:DExp_3MN} and Appendix \ref{app:expdisc}). As already remarked, the numerical evaluation of the potential is the most time consuming part of the construction of a model. For this reason JASMINE2 also contains a continuously updated library of analytical
density-potential pairs available in the literature \citep[and in some cases also based on homoeoidal expansion, see e.g.][]{ciot-bertin05}, so that one can choose between the numerical computation of the potential and (when available) the use of the analytical potential.

\subsubsection{The Post-Processing}

As described in the previous Section, for a given multi-component
model of assigned $\trhoi$, $\trhoh$, and with a central BH, the
Potential \& Jeans Solver gives the solution $\tsigmij^2$ and $\tDij$
of the scaled form of equations \eqref{eq:JEsij}. These solutions are
then combined in PP, with the assignment of the mass ratios $\RMj$, so
that the solution $\sigmi^2$ and $\Di$ of equations \eqref{eq:JEs} for
$\rhoi$ is obtained, according to equations \eqref{eq:pzi_deltai_sum}
and \eqref{eq:kin_tilde}.  At this stage, as discussed in Section
\ref{sec:satoh}, the PP performs a positivity check of $\tDi$ and
$\tDi+\tsigmi^2$: in case of negativity of the last quantity, a new
choice of the weights $\RMj$ is made, until positivity is reached. If
positivity cannot be obtained for acceptable choices of $\RMj$, then
the multi-component model is discarded as unphysical.

Once the mass weights are assigned and the positivity check is passed,
the PP requires the parameter $k_i$ for the kinematical decomposition,
that gives the scaled azimuthal velocity fields $\vphii$ and
$\sigphii$. In full generality, we define each decomposition parameter
as
\begin{equation}
k_i(R, z)=\lambda_i \delta_i(R,z), 
\label{eq:kfacti}
\end{equation}
where $\lambda_i$ is a constant weight, and $\delta_i(R,z)$ is a
position-dependent function; the standard Satoh parameter is obtained
with $\delta_i = 1$ and $\lambda_i= k_i$.  The benefit of this
factorisation is due to the fact that the projection formula of
$\vphii$ in equation \eqref{eq:sigma_vlos} for a given $\delta_i(R,z)$
scales with $\lambda_i$, so that we can set the value of $\lambda_i$
{\it after} having computed the projection integral. As projections
represent the second most time-consuming step, the possibility to
choose (and change) $\lambda_i$ after projections is a significant
advantage. Note that, at variance with what happens for the fields
$\tsigmi^2$, $\tDi$, ${\tilde\upsilon}_{\varphi i}^2$ and
$\tilde\sigma_{\varphi i}^2$, the mass weights $\RMj$ enter the
expression of ${\tilde\upsilon}_{\varphi i}$ under a square root (see
equations \ref{eq:satoh_i} and \ref{eq:g-satoh_i}). This implies that
the $\RMj$ must be chosen before calculating the projections that use
$\vphii$\footnote{ The \textit{effective radius} $\Reff$ of the total
  stellar distribution is obviously another important quantity that
  cannot be obtained as a linear combination of the effective radii of
  the stellar components, and it can only be computed after the choice
  of the weights $\RMi$ and $\mtoli$. }. In other words, the
possibility to modify the values of $\RMj$ in PP, allowed by the
''$ij$-decomposition'', ends with the computation of the scaled fields
in equations \eqref{eq:kin_tilde}.

Once we have obtained the intrinsic and projected fields of each
$\rhoi$, the last steps are to combine them to calculate the
\textit{total} (mass- and luminosity-weighted) intrinsic and projected
fields of $\rho_*$ (respectively from equations
\ref{eq:pz_Delta_rhvphi_sum}, \ref{eq:pphi_sum}, and equations
\ref{eq:sigma_vlos}, \ref{eq:sigmalos}, \ref{eq:sigmavlosi_sumi}), and
finally to choose the physical scales $\Mstar$ and $\rstar$.

\subsection{Summary}

Summarising, a {\it family} of multi-component galaxy models is
defined by the choice of $N$ scaled stellar density components
$\trhoi$, a scaled DM halo $\trhoh$, and a central BH. The Potential
and Jeans Solver computes the associated scaled potentials $\tphij$,
and then solves the $N\times (N+2)$ Jeans equations \eqref{eq:JEsij}
in their scaled form. In the subsequent PP, specific values of the
mass ratios $\RMi$, $\RMh$, $\RMbh$, of the mass-to-light ratios
$\mtoli$, and of the kinematical decompositions with the parameters
$k_i$, are fixed, thus defining a specific model in the same
family. The solution of the Jeans equations for the total density
distribution is recovered as (mass- or luminosity-) weighted sums of
the scaled solutions, and the projections along a given line-of-sight
are performed. The values of the total stellar mass $\Mstar$, and of
its scale-length $\rstar$, complete the construction of the model.

There are at least two significant advantages in this procedure, when
compared with a straightforward integration of the Jeans equations for
a multi-component galaxy model. First, the gravitational potentials of
each stellar component and of the DM halo need not to be recalculated
every time the weights are changed in PP; thus the run of the most
time expensive part of is required just once for all the models in the
same family. Second, the possibility to choose the weight parameters
in PP allows for a fast exploration of the parameter space (that, for
multi-component models, can be very large). Qualitatively, the
$N\times (N+2)$ set of the $ij$-th scaled solutions of the Jeans
equations for each $i$-th density component in each $j$-th potential
component, can be interpreted as {\it basis vectors} that are
successively linearly combined with different weights, to obtain a
specific solution belonging to a family of multi-component models.

As a final remark, note that the procedure described so far can be extended to more general velocity decompositions. For example, it is straightforward to insert in it the \citet{capp08} orbital anisotropy, where $\sigma_{Ri}=b_i \, \sigma_{zi}$, with $b_i$ a constant parameter that can be different for each stellar component, and the underlying $i$-th DF depends on 3 integrals of motion.

\section{Four illustrative multi-component models} \label{sec:app}

In order to illustrate the new features and potentialities of our
procedure, as implemented in JASMINE2, we firstly describe in some detail the building of three multi-component galaxy models. All three models are made of two stellar distributions, to which a DM halo with a spherical Navarro-Frenk-White profile \citep[][hereafter NFW]{NFW} and
a central supermassive BH are added. In the first model (hereafter JJE) the {\it total} spherical stellar profile and an ellipsoidal stellar component, both with a \citet{jaffe83} profile, are assigned; if the dark mass is set to zero, this model reduces to the JJe models of \citetalias{cmpz21}. The second model (hereafter JHD) consists of an ellipsoidal Jaffe stellar density distribution, that represents a light stellar halo, coupled with a heavy Miyamoto-Nagai stellar disc \citep[hereafter MN,][]{MN}. In the third model (hereafter JLD), the
ellipsoidal Jaffe component dominates, while a small MN inner disc is counter-rotating. These three models, are intended to represent features observed in real galaxies, but they are not designed to reproduce specific objects. Finally, we illustrate the comparison  between an exponential disc and its representation via the sum of three MN discs, as proposed by \citet{smith2015}.

\subsection{The JJE models}
\label{sec:JJE}

JJE models are a natural generalisation of JJe models presented in
\citetalias{cmpz21}: as these latter describe quite well real
elliptical galaxies, and several of their dynamical properties can be
expressed in analytical form, they also represent an obvious test for JASMINE2.

To better appreciate the properties of JJE models, we recall the main
properties (and limitations) of JJe models. These are constructed by
assigning a {\it total} density $\rho_*$ following the axisymmetric
ellipsoidal generalisation of the Jaffe model, and another
axisymmetric ellipsoidal Jaffe distribution $\rhof$, with different
flattening, scale-length and total mass; in \citetalias{cmpz21} the
density distribution $\rho_* - \rhof$ ($=\rhos$ in the current
notation), is interpreted as a DM halo; finally, a central BH is added
to the system. The analytical conditions on $\rhof$ to guarantee the
positivity of $\rhos$ are given, and then the Jeans equations for
$\rhof$ are solved in analytical closed form, by using homoeoidal
expansion, truncated at the linear order in the flattenings of
$\rho_*$ and $\rhof$. Albeit several properties of JJe models can be
expressed in analytical form \citep[making these models quite useful
in numerical simulations of gas flows in galaxies, see
e.g.][]{gan_ciot19, gan_choi19}, a few important shortcomings still
affect them: i) the Jeans equations for $\rhof$ are integrated in the
homoeoidal expansion limit, retaining only linear terms in the
flattenings, and they have not been studied for the difference
component $\rhos$; ii) projected kinematical fields of $\rhof$ can be
obtained in analytical form only as asymptotic formulae at the center
and at large radii. JASMINE2 is then the obvious tool to address the
two points above.

Here we generalise the JJe models to JJE models, by considering for
the total $\rho_*$ an ellipsoidal Jaffe profile, of total mass
$\Mstar$, scale-length \footnote{In the spherical limit, and in the
  assumption of constant mass-to-light ratio, the Jaffe scale radius
  $\rstar$ is related to the effective radius by
  $R_\rme \simeq 0.75 \, \rstar$.} $\xi$, and flattening $q$:
\begin{equation}
 \rhostar(R,z) = {\rhon\xi\over q \, m^2 (\xi+m)^2}, \quad m^2 =
 \tilde R^2+{\tilde z^2\over q^2}. 
\label{eq:rho_jaffe}
\end{equation}
Therefore, at variance with JJe models, in JJE models the total Jaffe
mass distribution is purely stellar. We then consider another
ellipsoidal Jaffe density profile, of total mass $\Msone=\RMo \Mstar$,
scale-length $r_{*1}=\xi_1 \rstar$, and flattening $q_1$:
\begin{equation} 
  \rhof(R,z) = {\rhon \RMo\xi_1\over q_1 m_1^2 (\xi_1+m_1)^2},\quad
  m_1^2 = \tilde R^2+{\tilde z^2\over q_1^2}.
 \label{eq:rho1_jaffe}
\end{equation}
The second stellar component is then defined as
\begin{equation}
  \rhos(R,z) = \rhostar(R,z) - \rhof(R,z), 
\label{eq:rho2_diff}
\end{equation}
with $\Mstwo = \Mstar - \Msone = (1-\RMo)\,\Mstar=\RMt\,\Mstar$, in
agreement with equation \eqref{eq:sumRi}. Notice that $\rhos$ is not
an ellipsoid, unless $q_1=q$, and even in this case $\rhos$ is not a
Jaffe ellipsoid, unless $\xi_1=\xi$. As extensively discussed in
\citetalias{cmpz21}, $\rhos$ could be negative somewhere (and so
unphysical) for some choices of $\RMo$, $\xi_1$ and $q_1$. Remarkably,
the conditions required to assure $\rhos\geq 0$ can be expressed as
analytical (and simple) inequalities, as shown in Appendix
\ref{app:pos_cond}.

The stellar distribution $\rho_*$ is embedded in a NFW DM halo
(spherically symmetric for simplicity), of mass
$\Mh (\rt)=\RMh \Mstar$ enclosed within a truncation radius $\rt$,
scale-length $\rh=\xi_\rmh \rstar$, and concentration
$c\equiv \rt/\rh$:
\begin{equation}
  \rhoh(r) ={\rhon\RMh\over f(c) s (\xi_\rmh +s)^2},\quad
\phih(r) = - \phin\RMh{\ln (1+ s/\xi_\rmh)\over  f(c)s}\,
\label{eq:NFW}
\end{equation}
where  $s\equiv r/\rstar$, and $f(c) = \ln(1+c)-c /(1+c)$. We complete the model with a central BH of mass 
$\MBH=\RMbh \Mstar$.

Summarising, JJE models are determined, besides the total stellar mass
and scale-length, $\Mstar$ and $\rstar$, by the two parameters $\xi$
and $q$ for $\rho_*$, the five parameters $q_1$, $\xi_1$, $\RMo$,
$\MLo$, $k_1$ for $\rhof$, the two parameters $\MLt$, $k_2$ for
$\rhos$, the three DM parameters $\xi_\rmh$, $c$, $\RMh$, and the BH
mass weight $\RMbh$ (see Table \ref{tab:model_parameters} for a
specific JJE model). JASMINE2 further generalises JJE models, with the
addition of a second DM component given by a shallow and very extended
quasi-isothermal halo, as useful in simulations of gas flows in
galaxies residing in groups or clusters (\citetalias{ciot_gan21} et al. in
preparation).

%%%%%%%%%%%%%%%%%%%%%%%%%%%%%%%%%%%%%%%%%%%%%%%%%%%%%%%
\begin{table}
 \centering 
 \begin{tabular}{ccc}
  \toprule\toprule 
  Model & $\rho_{*1}$ & $\rho_{*2}$ \\
  \midrule\midrule 
  JJE & Jaffe & $\rhostar(\xi=1,q=1) -\rho_{*1}$ \\
  & $\xi_1=0.1$ & --- \\
  & $q_1=0.8$ & --- \\
%  \cmidrule(lr){2-3}
  & $\RMo=0.04$ & $\RMt=0.96$ \\
  & $\MLo=2$ & $\MLt=6$ \\
  & $k_1=0.5$ & $k_2=0.2$\\
 \midrule 
  JHD & Jaffe & MN \\
  & $\xi_1=1$ & $\tilde b=0.1$ \\
  & $q_1=0.8$ & $q_2 =10$ \\
%  \cmidrule(lr){2-3}
  & $\RMo=0.3$ & $\RMt=0.7$ \\
  & $\MLo=6$ & $\MLt=2$ \\
  & $k_1=0.5$ & $k_2=0.8$\\
 \midrule 
  JLD & Jaffe & MN \\
  & $\xi_1=1$ & $\tilde b=0.01$ \\
  & $q_1=0.8$ & $q_2=10$ \\
%  \cmidrule(lr){2-3}
  & $\RMo=0.96$ & $\RMt=0.04$ \\
  & $\MLo=6$ & $\MLt=2$ \\
  & $k_1=0.5$ & $k_2(R,z)$  \\
  \bottomrule\bottomrule 
 \end{tabular}
 \caption{The parameters for the stellar components of the
   illustrative JJE, JHD and JLD models (Sections \ref{sec:JJE} and
   \ref{sec:JHD&JLD}). In the JJE model, the component $\rho_{*2}$ is
   obtained as {\it difference} between a total spherical ($q=1$)
   Jaffe profile $\rhostar$, with scale-length $\xi=1$, and a small
   and light Jaffe ellipsoidal component $\rho_{*1}$ (as done for JJe
   models in \citetalias{cmpz21}). The standard Satoh
   $k$-decomposition in equation \eqref{eq:satoh_i} for $\rho_{*1}$,
   and the generalised $k$-decomposition in equation
   \eqref{eq:g-satoh_i} for $\rho_{*2}$, are adopted. In the JHD
   model, an ellipsoidal Jaffe distribution is coupled to a massive
   and quite flat ($q_2=a/b=10$) MN disc; in both components a
   generalised $k$-decomposition is adopted. In the JLD model, the
   ellipsoidal Jaffe component has the same flattening and size as in
   the JHD model, but the disc is significantly smaller, and
   counter-rotates in the inner regions, with the position-dependent
   Satoh parameter in equation \eqref{eq:kcontr}, while a constant
   Satoh parameter is applied to the Jaffe component. In all models,
   the DM halo has a spherical NFW profile with $\xi_\rmh=2.6$,
   $c=10$, $\RMh=20$, and the BH is defined by $\RMbh=0.002$.}
 \label{tab:model_parameters}
\end{table}
%%%%%%%%%%%%%%%%%%%%%%%%%%%%%%%%%%%%%%%%%%%%%%%%%%%%%%%

\subsubsection{Tests}
\label{sec:tests_JJE}

We can use JASMINE2 in two different tests: we can give in input the homoeoidal expansion of the density-potential pairs, truncated at the linear order, and compare the numerical solution with that of \citetalias{cmpz21}, to check the importance of quadratic flattening terms in the solution of the Jeans equations for JJe models; and we can give in input the \textit{true} ellipsoidal model, to check how well the homoeoidal expansion reproduces its internal dynamics. At
the same time, the previous tests allow for an accuracy check of JASMINE2.

In the first test, we feed JASMINE2 with the homoeoidal
expansion for $\rhostar$ and $\rhof$, and we compare the numerical
results of integration of the Jeans equations with the analytical
results: we obtain excellent agreement for all the kinematical fields,
better than a fraction of percent over the whole numerical grid,
ranging from $\approx 10^{-5}$ to $\approx 70$ (in units of
$\rstar$). The results tend to be slightly more discrepant at
increasing flattenings, as expected, since the analytical results in
\citetalias{cmpz21} are limited to the {\it linear} order in the
flattenings, while JASMINE2 takes automatically into account also the
{\it second} order terms when integrating the Jeans equations (due to
the product between the density and the gradient of the potential). By
increasing the numerical resolution in the central regions, and moving
to smaller and smaller distances from the center, we also verify
that the asymptotic formulae for the projected kinematical fields are
also perfectly recovered numerically. This first tests adds
confidence that JASMINE2 is working properly and with high accuracy,
but also provides a further support that the (quite cumbersome)
analytical formulae in \citetalias{cmpz21} are actually correct
\footnote{ These tests are similar in the approach to those
  already performed with JASMINE by using the analytical results of
  \citet{smet_pos15}: we recall that the numerical integration of the
  potential in JASMINE2 uses the same routines of JASMINE.}.

In a second test, we compare the numerical results of JASMINE2
for the {\it true} ellipsoidal JJe models with the analytical results
in \citetalias{cmpz21} obtained from homoeoidal expansion, therefore
moving beyond the effect of second order approximation in the
flattenings explored in the first test. We find that for
relatively small flattenings (corresponding to E2-E3
galaxies) the homoeoidal expansion truncated at the linear order
provides quite good results, even when compared with true ellipsoidal
models, with the most significant discrepancy in the intermediate
regions.

%%%%%%%%%%%%%%%%%%%%%%%%%%%%%%%%%%%%%%%%%%%%%%%%%%%%%%%%%%%%%%%%%%%%%
\begin{figure*} % figure* to use both two columns 
 \centering 
 %\hspace{-.1cm}
 \subfloat{\includegraphics[width=.98\textwidth]{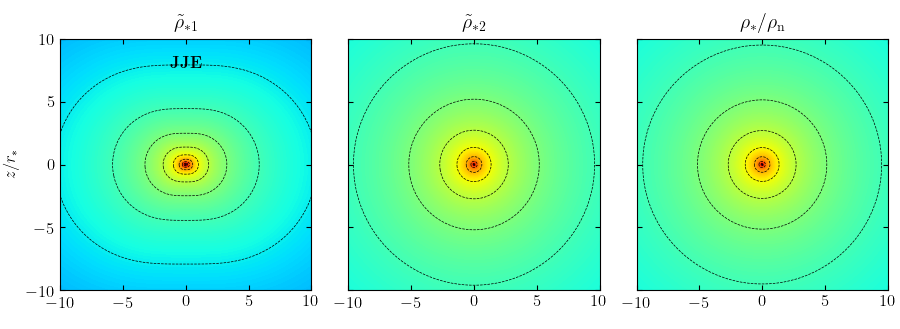}} \\ \vspace{-.75cm} %\hspace{-.2cm}
 \subfloat{\includegraphics[width=.98\textwidth]{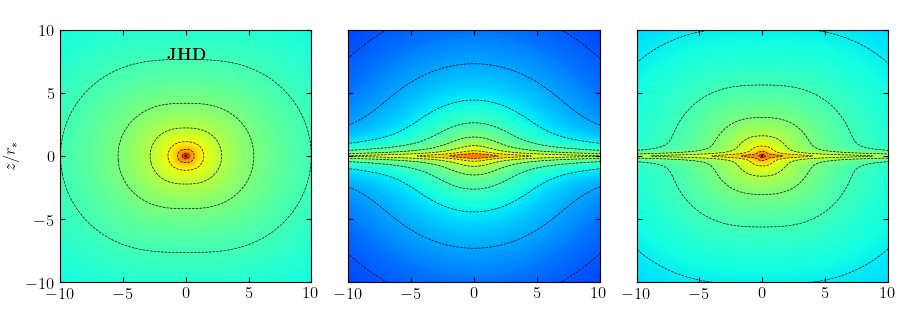}} \\ \vspace{-.5cm}
 \subfloat{\includegraphics[width=\textwidth]{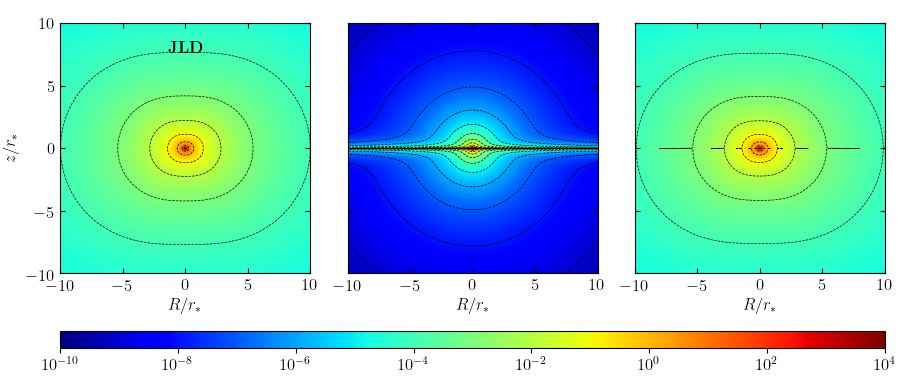}}
 \caption{The scaled stellar distributions $\trhof$, 
   $\trhos$, and the dimensionless total stellar distribution 
   $\trhostar=\rho_*/\rhon$, of the three models of Table \ref{tab:model_parameters}. The dotted contours show the isodensities, with values spaced by 1 dex.}
 \label{fig:rho}
\end{figure*}
%%%%%%%%%%%%%%%%%%%%%%%%%%%%%%%%%%%%%%%%%%%%%%%%%%%%%%%%%%%%%%%%%%%%%

%%%%%%%%%%%%%%%%%%%%%%%%%%%%%%%%%%%%%%%%%%%%%%%%%%%%%%%%%%%%%%%%%%%%%
\begin{figure*} % figure* to use both two columns 
 \centering 
 \subfloat{\includegraphics[width=\textwidth]{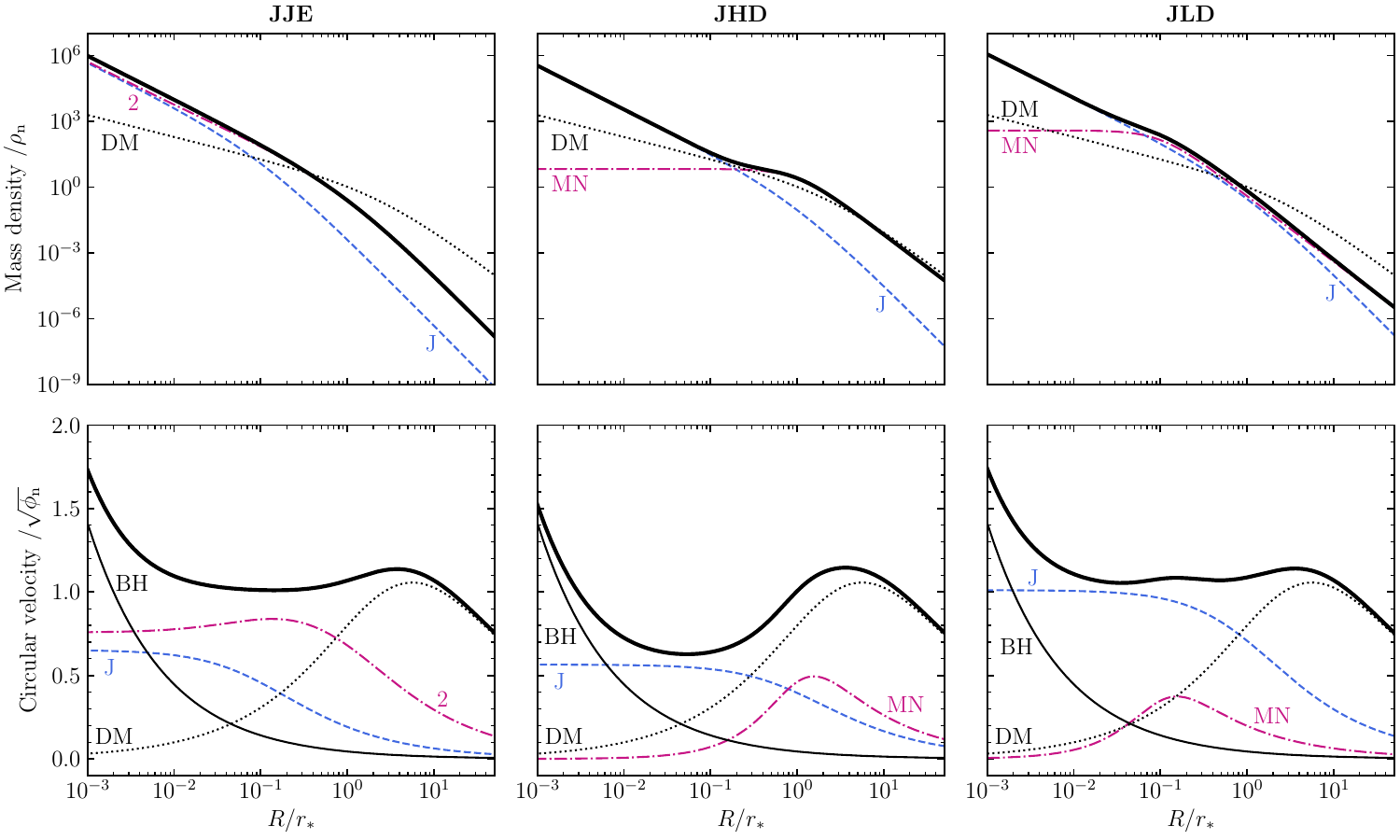}}
  \caption{Radial profiles in the equatorial plane ($\tilde z=0$) of
   the mass densities $\RMo\trhof$ (dashed blue), $\RMt\trhos$
   (dotted-dashed magenta), $\trhostar$ (heavy solid), and
   $\RMh\tilde\rho_\rmh$ (dotted), normalised to $\rhon$ (top row),
   for the three models of Table \ref{tab:model_parameters}. In the
   bottom row, we show the corresponding contributions to the total
   circular velocity (heavy solid) in the equatorial plane of the mass
   components, with the additional contribution of the central BH
   (solid), all normalised to $\sqrt{\phin}$.}
 \label{fig:rho_vc_z0}
\end{figure*}
%%%%%%%%%%%%%%%%%%%%%%%%%%%%%%%%%%%%%%%%%%%%%%%%%%%%%%%%%%%%%%%%%%%%%

%%%%%%%%%%%%%%%%%%%%%%%%%%%%%%%%%%%%%%%%%%%%%%%%%%%%%%%%%%%%%%%%%%%%%
\begin{figure*} % figure* to use both two columns 
 \centering 
 \subfloat{\includegraphics[width=\textwidth]{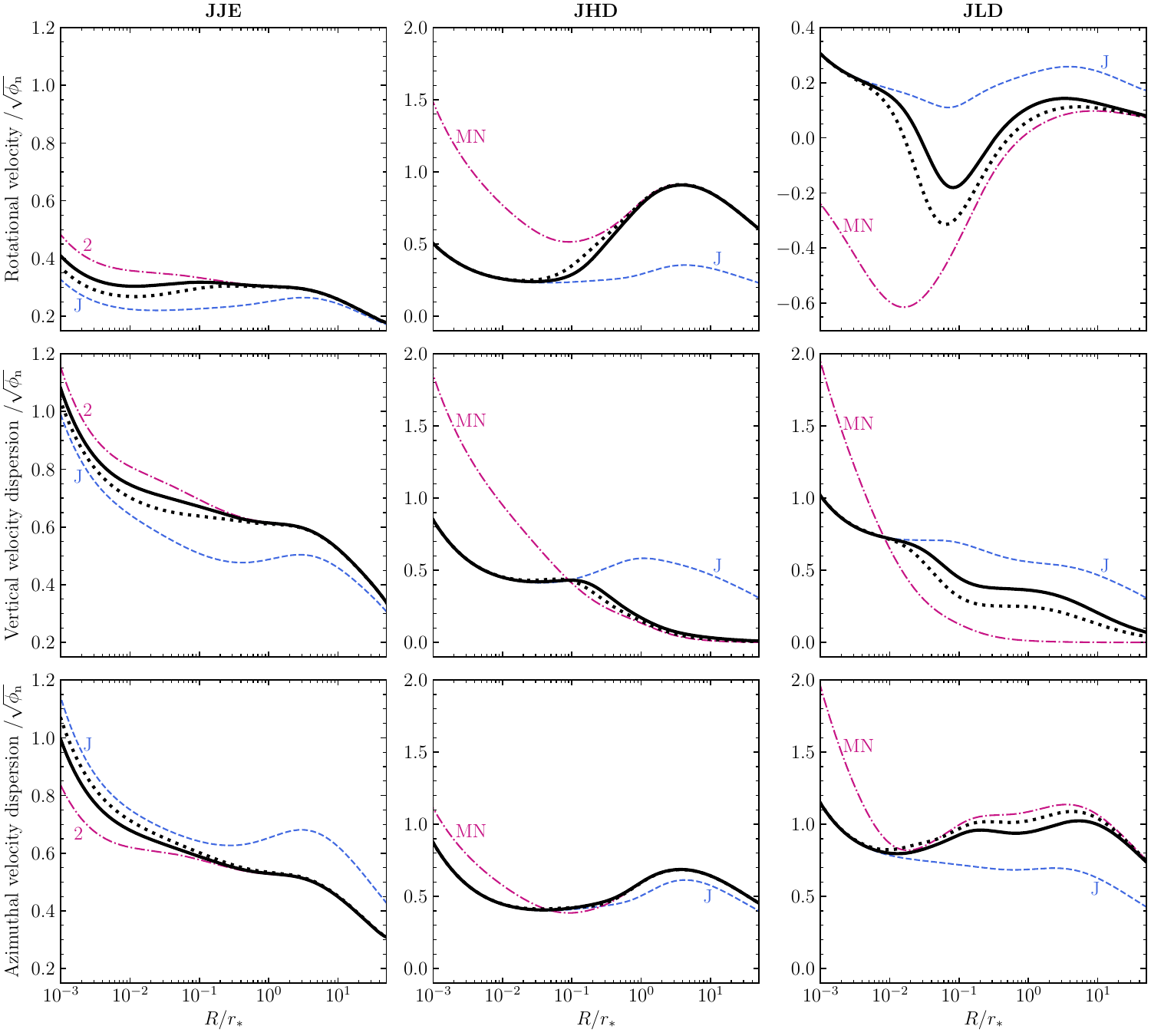}}
 \caption{Radial profiles in the equatorial plane ($\tilde z=0$) of
   the rotational velocities (top row), vertical velocity dispersions
   (middle row), and azimuthal velocity dispersions (bottom row),
   normalised to $\sqrt{\phin}$, for the three models of Table
   \ref{tab:model_parameters}. Each panel shows the total
   mass-weighted (heavy solid) and luminosity-weighted (heavy dotted)
   fields, together with the corresponding fields of $\rho_{*1}$
   (dashed blue) and $\rho_{*2}$ (dotted-dashed magenta). Notice the
   different values on the vertical scales of the first column (JJE
   model) and of the top right panel showing the counter-rotation (JLD
   model).}
 \label{fig:vel_z0}
\end{figure*}
%%%%%%%%%%%%%%%%%%%%%%%%%%%%%%%%%%%%%%%%%%%%%%%%%%%%%%%%%%%%%%%%%%%%%

%%%%%%%%%%%%%%%%%%%%%%%%%%%%%%%%%%%%%%%%%%%%%%%%%%%%%%%%%%%%%%%%%%%%%
\begin{figure*} % figure* to use both two columns 
 \centering 
 \subfloat{\includegraphics[width=.8\textwidth]{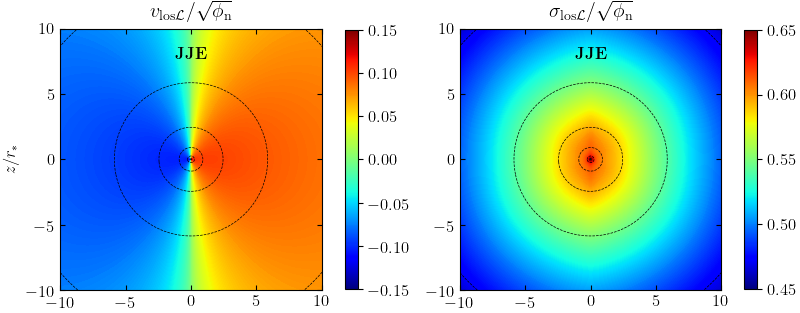}} \\ \vspace{-.45cm}
 \subfloat{\includegraphics[width=.8\textwidth]{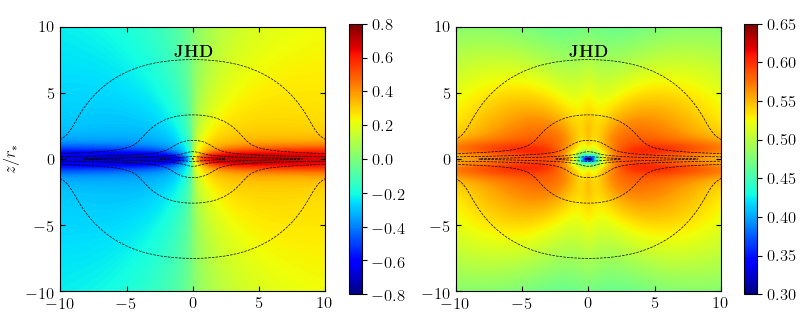}} \\ \vspace{-.45cm}
 \subfloat{\includegraphics[width=.8\textwidth]{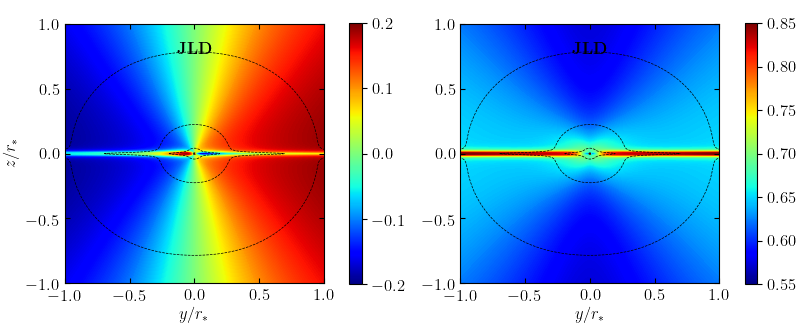}}
 \caption{Edge-on projected luminosity-weighted rotational velocity
   $\upsilon_{\mathrm{los}\mathscr{L}}$ (left), and velocity
   dispersion $\sigma_{\mathrm{los}\mathscr{L}}$ (right), normalised
   to ${\sqrt{\phin}}$, for the three models of Table
   \ref{tab:model_parameters}. Notice the different ranges of values
   on the colorbars. For the JLD model, the region shown is limited to
   $r_*$ to appreciate the central features, in particular the inner
   counter-rotating disc. The dotted contours show the galaxy
   isophotes  with values spaced by 1 dex.}
 \label{fig:losvel}
\end{figure*}
%%%%%%%%%%%%%%%%%%%%%%%%%%%%%%%%%%%%%%%%%%%%%%%%%%%%%%%%%%%%%%%%%%%%%

\subsubsection{Results for a JJE model}
\label{sec:results_JJE}

We move now to illustrate the main properties of a specific JJE model
(see Table \ref{tab:model_parameters}). The total stellar distribution
$\rho_*$ has a spherical Jaffe profile obtained from equation
\eqref{eq:rho_jaffe} with $\xi=1$ and $q=1$; this quite artificial
case allows us to discuss some subtleties that can occur to the
kinematical decomposition in multi-component systems. The stellar
component $\rhof$ is obtained from equation \eqref{eq:rho1_jaffe} with
$\xi_1=0.1$, $q_1=0.8$, $\RMo$ is $0.04$, and $\MLo=2$, i.e. it is a
quite small ellipsoidal distribution at the center of the galaxy; note
that, from equation \eqref{eq:pos_cond_final}, the maximum possible
value of $\RMo$ is $0.08$.  The component $\rhos$ accounts for the
remaining $96\%$ of the total stellar mass of the galaxy, and
$\MLt=6$, so that $\rho_*$ could represent an elliptical galaxy with a
central and younger stellar system. We add the spherical NFW DM halo,
given by equation \eqref{eq:NFW}, with $\xi_\rmh=2.6$, $c=10$, and
$\RMh=20$, so that the DM mass inside $\Reff$ (see Footnote 3) is
$\approx 0.45$ of the total mass. Finally, in agreement with BH-galaxy
scaling relations \citep[see e.g.][]{korm-ho13}, the mass of the
central BH is fixed to $\RMbh=0.002$.

In the three top panels of Figure \ref{fig:rho}, we show the density
distribution of the scaled components $\trhof$ and $\trhos$, and of
the total stellar density $\trhostar=\rho_*/\rhon$. Being this last
spherical, and $\trhof$ oblate, $\trhos$ in its central regions is
slightly prolate, and this will affect its kinematical fields, as
anticipated in Section \ref{sec:satoh}. Additional information on the
model structure is provided in the first column of Figure
\ref{fig:rho_vc_z0}: the top panel shows the radial profiles in the
equatorial plane of $\RMo\trhof$, $\RMt\trhos$, $\trhostar$, and
$\RMh\trhoh$. The total $\rhostar$ is almost coincident with $\rhos$,
except for the central regions, where $\rhof$ and $\rhos$ are
comparable. The DM density $\rhoh$ overcomes $\rhostar$ outside
$\approx 0.5\,\Reff$. The bottom panel shows the radial profiles in
the equatorial plane of the contributions to the circular velocity due
to the various mass components: the BH contribution is dominant in the
inner regions, the DM in the outer regions, while at intermediate
distances from the centre the resulting circular velocity is quite
flat.

Similar trends can be seen in the radial profiles of the velocity
fields in the equatorial plane of Figure \ref{fig:vel_z0}, where, in
the first column from top to bottom, we show the rotational velocity,
the vertical velocity dispersion, and the azimuthal velocity
dispersion, of $\rhof$ and $\rhos$, and the total mass-weighted and
luminosity-weighted fields. Note that in the three panels the vertical
scale is the same, and the resulting system is clearly a slow
rotator. This JJE model offers the opportunity to apply the
generalised $k$-decomposition in equation \eqref{eq:g-satoh_i};
because the field $\Delta_2$, associated to the slightly prolate
$\rhos$, is negative in the central regions. As discussed in Section
\ref{sec:satoh}, we verified that $\barvphi_2$ is nowhere negative,
and then we adopted the generalised decomposition, with a quite small
$k_2=0.2$. The field $\Delta_1$ instead is everywhere positive, as
expected, and so we adopted the standard Satoh formula with
$k_1=0.5$. In the velocity profiles, the effect of the central BH is
clearly visible; for example, the velocity dispersion profile of a
Jaffe model with $\RMbh=0$ would be constant in the central
regions. Notice also how the velocity profiles, outside
$\approx \Reff$, are almost coincident with the profiles of the more
massive component $\rhos$, in both the mass-weighted and
luminosity-weighted cases; this is not surprising, because in these
regions $\rhostar$ coincides with $\rhos$ (see Figures \ref{fig:rho}
and \ref{fig:rho_vc_z0}). The situation is different in the inner
regions, where $\rhof$ and $\rhos$ are comparable: here the total
velocities have intermediate values, with the luminosity-weighted
profiles closer to the profiles of $\rhof$ with the smaller
mass-to-light ratio.

As an illustration of the projection procedure, in the first row of
Figure \ref{fig:losvel}, we show the EO projected luminosity-weighted
fields $\vlosL$ and $\siglosL$, with the superimposed dotted contours
representing the galaxy isophotes of the surface brightness $I_*$. The
slow rotation of the model is apparent from the colorbar values,
indeed $\vlosL$ is everywhere lower than $\siglosL$. A curious feature
is the slightly vertically elongated shape of $\siglosL$: this is
\textit{not} due to the prolate shape of $\rhos$ in the central
regions, but it is an effect of the generalised $k$-decomposition,
coupled with the fact that $\Delta_2$ is almost null in the external
regions, and so here $\upsilon_{\varphi 2} \sim k_2 \sigma_2$, as
introduced in Section \ref{sec:satoh}. For example, an increase in
$k_2$ would lead to an increase of the rotation in the external
regions, with correspondent decrease of $\siglosL$, and with the net
result of a more elongation of $\siglosL$ in the central regions.

\subsection{Ellipsoidal models with an embedded stellar disc} 
\label{sec:JHD&JLD}

JHD and JLD models consist of a stellar profile $\rhof$ given again by
the ellipsoidal Jaffe model in equation \eqref{eq:rho1_jaffe}, coupled
with a stellar MN disc $\rhos$, of total mass $\Mstwo=\RMt\Mstar$, and
scale-lengths $a=\tilda\rstar$, $b=\tildb\rstar$:
\begin{equation}
 \rhos(R,z) = \rhon \RMt \tildb^2 
{\tilda\tildR^2 + (\zeta +2\sqrt{\tildz^2+\tildb^2})\zeta^2\over 
  (\tildR^2+\zeta^2)^{5/2}(\tildz^2+\tildb^2)^{3/2}},
\label{eq:MNrho}
\end{equation}
\begin{equation}
 \phi_{*2}(R,z) = - {\phin \RMt\over\sqrt{\tildR^2 + \zeta^2}},
   \quad 
   \zeta=\tilda + \sqrt{\tildz^2 + \tildb^2},
   \label{eq:MNphi}
\end{equation}
where $\RMt=1-\RMo$ from equation \eqref{eq:sumRi}.  For $a =0$ the MN
disc reduces to the \citet{plummer11} sphere, and for $b=0$ to the
razor-thin \citet{kuzmin56} disc; in the following, we indicate with
$q_2=a/b$ the disc flattening parameter. As in JJE models, we add the
spherical NFW halo in equation \eqref{eq:NFW}, and a central BH, so
that the resulting multi-component models are completely determined
once the values of $\xi_1$, $q_1$, $\RMo$, $\MLo$, $k_1$ for $\rhof$,
$\tildb$, $q_2$, $\MLt$, $k_2$ for $\rhos$, $\xi_\rmh$, $c$, $\RMh$
for $\rhoh$, and $\RMbh$ for the BH, are assigned, in addition to the
total stellar mass $\Mstar$ and the scale length $\rstar$ (see Table
\ref{tab:model_parameters}). In Section \ref{sec:results_JHD} we
consider the case of a dominant MN disc, when the ellipsoidal Jaffe
component can be interpreted as the stellar halo of a disc galaxy, and
in Section \ref{sec:results_JLD} the case of a small and
counter-rotating stellar disc at the center of a dominant stellar
spheroid, as sometimes observed in real ETGs (e.g. \citealt{morelli04,
  kraj15, mitzkus17}; see also \citealt{capp-rev16}). The parameters
of the DM halo and of the central BH are the same as in the JJE model.

\subsubsection{Results for the JHD model}
\label{sec:results_JHD}

In the ''Jaffe - Heavy Disc'' JHD model (see Table
\ref{tab:model_parameters}), the ellipsoidal Jaffe stellar halo
$\rhof$ is characterised by a scale-length $\xi_1=1$, a flattening
$q_1=0.8$, a stellar mass fraction of $30\%$ (i.e. $\RMo=0.3$), and a
mass-to-light ratio $\MLo=6$. The dominant MN disc $\rhos$
($\RMt=0.7$) is quite flat ($q_2=10$), with $\tilde b=0.1$, and a
lower $\MLt=2$.

In the central row of Figure \ref{fig:rho}, the scaled density
distributions $\trhof$, $\trhos$, and $\trhostar$, are shown. The
resulting isodensity contours of $\rho_*$ would be classified as
''discy'' near the equatorial plane, and as ''boxy'' at large distance
from the plane. The radial profiles of the density distributions
(including the DM), in the equatorial plane, are shown in Figure
\ref{fig:rho_vc_z0}. It is apparent how, inside $\approx 0.1\, r_*$
the Jaffe halo dominates, around $r_*$ the MN disc dominates, and
$\rhoh$ overcomes the total $\rho_*$ outside $\approx 10\, r_*$. Note
that, even if $\RMo<\RMt$, $\rhof$ dominates the total density in the
central regions, due to the cuspy profile of the Jaffe density
compared with the flat core of the MN density. The density
decomposition reflects on the circular velocity profiles in the bottom
panel of the same Figure: the total $\upsilon_{\rm c}$ at small radii
is totally dominated by the BH, and at large radii by the DM halo;
while the ''bump'' around $3\,r_*$ is due to the MN and the DM
potentials.

The radial profiles in the equatorial plane of the velocity fields,
obtained from the Jeans equations, are shown in the middle column of
Figure \ref{fig:vel_z0}, where from top to bottom the total mass- and
luminosity-weighted rotational velocity, vertical velocity dispersion,
and azimuthal velocity dispersion, are plotted together with the
corresponding quantities for each stellar component separately. For
the adopted values of the parameters in Table
\ref{tab:model_parameters}, $\Delta_1$ turns out to be negative in a
quite central region, while $\Delta_1+\sigma_1^2$ is everywhere
positive; we decided to apply the generalised $k$-decomposition in
equation \eqref{eq:g-satoh_i} to both stellar components, with
$k_1=0.5$ and $k_2=0.8$. The total velocity profiles, in the central
regions, are completely determined by the Jaffe profile, because here
$\rhof > \rhos$, compensating also for the higher $\MLo$; in the
external regions, instead, the total profiles are dominated by the MN
disc. Furthermore, $\vphi$ stays well below $\upsilon_{\rm c}$ both in
the inner and outer regions (see $\upsilon_{\rm c}$ in Figure
\ref{fig:rho_vc_z0}), as a clear manifestation of \textit{asymmetric
  drift} in the equatorial plane (e.g. \citetalias{b&t08}). Note that
$\sigma_2$, associated with a flat density profile at the centre, is
much higher than $\sigma_1$, associated with $\rhof \sim R^{-2}$ in
the inner regions, as can be expected from the integration of the
vertical Jeans equation for a power law density distribution in the
gravitational field of a point-mass (i.e. the BH). In addition,
$\Delta_2$ of the MN model with the central BH vanishes at the centre
\citep[see e.g. Chapter 13 in][]{ciotti21}, thus in the generalised
$k$-decomposition, $\upsilon_{\varphi 2} \sim k_2 \sigma_2$ (at
variance with what would happen in the standard Satoh decomposition,
i.e. $\upsilon_{\varphi 2} = k_2 \sqrt{\Delta_2}$).

In the second row of Figure \ref{fig:losvel}, the luminosity-weighted
projected fields $\vlosL$ and $\siglosL$ are shown, and the high
rotation of the disc is clearly visible. The drop of $\vlosL$ inside
$r_*$ is due to a drop of the intrinsic rotational velocity (Figure
\ref{fig:vel_z0}). Also $\siglosL$ shows the highest values near the
equatorial plane, with a toroidal distribution around the centre, and
a drop inside $r_*$.

\subsubsection{Results for the JLD model} 
\label{sec:results_JLD}

At variance with the JHD model, in the ''Jaffe - Light Disc'' JLD
model (see Table \ref{tab:model_parameters}), the ellipsoidal Jaffe
distribution $\rhof$ accounts for almost the whole stellar mass of the
galaxy ($\RMo=0.96$), while its scale-length ($\xi_1=1$), flattening
($q_1=0.8$), and mass-to-light ratio ($\MLo=6)$ are unchanged. The
component $\rhos$ is a small MN disc, with $\tilde b=0.01$ and
$\RMt=0.04$, while $q_2=10$ and $\MLt=2$ are the same of the JHD
model.

The scaled density distributions, and the resulting total stellar
density, are shown in the three bottom panels of Figure \ref{fig:rho}:
$\trhof$ is (structurally) identical to that of the JHD model, while
$\trhos$ is much more concentrated, so that the total stellar density
is distributed in an extended halo with a very small disc. Indeed, the
disc is almost invisible in the last panel, and it would be apparent
only with a zoom in, as in Figure \ref{fig:losvel}. The last column of
Figure \ref{fig:rho_vc_z0} shows the radial profiles in the equatorial
plane of the density components, with their mass weights, and the
resulting decomposition of the galaxy circular velocity
profile. Notice that the central values of $\rhos$ are higher than
those in the JHD model, due to its smaller size, compensating for the
reduced mass. In the circular velocity plot, this reflects into a
larger contribution from the Jaffe component, and a smaller and inner
''bump'' of the MN component. As a result, $\upsilon_{\rm c}$ is
almost flat between $10^{-2}\,r_*$ and $10\,r_*$.

In the last column of Figure \ref{fig:vel_z0}, the radial profiles of
the velocity fields in the equatorial plane are shown. As in the
previous models, of course, the total luminosity-weighted profiles,
when distinguishable from the mass-weighted ones, are always closer to
the profiles of the component with the smaller mass-to-light
ratio. For the JLD model, both $\Delta_1$ and $\Delta_2$ are
everywhere positive, so we apply the standard Satoh decomposition. The
stellar halo is modeled as a slow rotator with $k_1=0.5$, while the
circumnuclear stellar disc as a faster and counter-rotating light
disc. In order to have counter-rotation limited to a central region,
we adopt a position-dependent Satoh parameter, defined as follows:
\begin{equation}
 k_2(R,z)=k_0+(k_{\infty}-k_0){r \over r+0.1 r_*}\,, \quad
 r=\sqrt{R^2+z^2}
 \label{eq:kcontr}
\end{equation}
(e.g. \citetalias{ciot_gan21} et al. in preparation; see also \citealt{negri14}
for an alternative parametrisation), with $k_0=-0.8$,
$k_{\infty}=0.1$, where the negative sign of $k_0$ assures the
counter-rotation of the disc, as can be seen in the top right panel of
Figure \ref{fig:vel_z0}. At very small radii (inside $10^{-2}\,r_*$),
the total rotational velocity is positive because the density is
dominated by the Jaffe component. We stress that the module of
$\upsilon_{\varphi 2}$ decreases towards the centre, at variance with
the JHD model, because now $\upsilon_{\varphi 2} = k_2 \Delta_2$, and
$\Delta_2 \rightarrow 0$, as explained in the previous Section. The
central total vertical velocity dispersion is higher than that of the
JHD model, even if the Jaffe component is structurally identical,
because of the higher $\RMo$ and of the more concentrated MN disc.

In the last row of Figure \ref{fig:losvel}, the los
luminosity-weighted velocities show clearly the effect of the inner
thin disc; the region shown is limited to $r_*$ to appreciate the
central features. In particular, in the $\vlosL$ distribution we have
counter-rotation at small radii (but not in the very centre). The disc
is also responsible for the highest values of the $\siglosL$ in the
equatorial plane, and the extended surrounding toroidal distribution
is also present, in analogy with the JHD model.

\subsection{Exponential discs and multi-MN decompositions}
\label{sec:DExp_3MN}

Exponential discs are the common choice for modeling disc
galaxies. Their gravitational potential can be constructed numerically by using the general formula based on complete elliptic integrals, or by using Bessel functions. The latter approach is particularly useful in case of factorised densities, such as
\begin{equation}
 \rhostar(R,z)=\rho_0{\rm e}^{-R/\Rd}V(|z|/h), 
 \label{eq:fact_exp_rho}
\end{equation}
where the function $V$ describes the vertical structure of the disc, and $\Rd$ and $h$ are respectively its scale-length and scale-height; the razor-thin exponential disc of central surface density $\Sigma_0$ is obtained for $V=\delta(z/h)$ and $\rho_0=\Sigma_0/ h$. Two natural generalisations of the infinitely thin exponential disc are obtained when $V$ is also an exponential function (double-exponential disc) or some negative power of the $\cosh$ function (''pseudo-isothermal'' exponential disc).
 
Unfortunately, the gravitational potential of these discs cannot be
obtained analytically; however, due to their relevance in the
construction of galaxy models, alternative models with analytical
potential have been proposed. In particular, the possibility to use
multi-component MN models to reproduce exponential discs, over some
finite radial range, has been explored for example by
\citet{smith2015} and \citet{rojas-nino2016} \citep[see
also][]{flynn96,ciot-pel96}. Such alternatives optimise the fit of the
density profile, and produce a good agreement with the circular velocity
profile of the exponential disc. Obviously, the superposition of MN
discs with their power-law radial decay at large radii (equation
\eqref{eq:MNrho}) cannot reproduce the exponential decay of equation
\eqref{eq:fact_exp_rho}. This forces to include at least one MN
density component with negative mass (or negative scale-length),
that can lead to a disc density distribution somewhere negative.
The use of a multi-component MN representation of an exponential disc
is motivated by the advantage of avoiding a time-consuming numerical
computation of its gravitational potential.
However, as we show in Appendix \ref{app:expdisc}, it is possible to
obtain the potential of factorised exponential discs in equation
\eqref{eq:fact_exp_rho} also with a very fast 1-dimensional
integration in terms of Bessel functions
(a method we implemented in JASMINE2).

As a last and natural application of our procedure, we extend the work
carried out by \citet{smith2015}  by constructing the solutions of the
Jeans equations for the double-exponential disc and for its everywhere
positive density representation in terms of three MN discs (hereafter
3MN). This 3MN decomposition is an ideal application of our modeling
procedure, in particular
because one MN component has negative mass, which gives the
opportunity to illustrate how the scaling scheme in Section
\ref{sec:scaling} works also with negative values of the mass ratios $\RMi$.

\subsubsection{Results for a double-exponential disc and its 3MN fit}
\label{sec:results_DExp_3MN}

We consider the single-component double-exponential model in equation
\eqref{eq:DExp_rho}, with mass $\RMd=M_{\rm d}/\Mstar=1$, scale-length
$\alpha=\Rd/\rstar=1$, and scale-height $\beta=h/\rstar=0.1$. For this
density, we build the everywhere positive 3MN fit, following Section
2.2 in \citet{smith2015}. Accordingly, the three MN components (in our
notation of equation \ref{eq:MNrho}) have the same scale-height
$\tilde b$, but different $\RM_i$ and
scale-length $\tilde{a}_i$ ($i=1,2,3$); in particular, from their Figure 5, we obtain $\tilde b=0.12$, and from their equation (7) the values of $\RM_i$ and $\tilde a_i$. The parameters for the double-exponential disc and for its 3MN fit are summarised in Table \ref{tab:DExp_3MN_parameters}.

\begin{figure} % figure* to use both two columns
 \centering 
 \vspace{-1.5cm}
 \subfloat{\includegraphics[width=.45\textwidth]{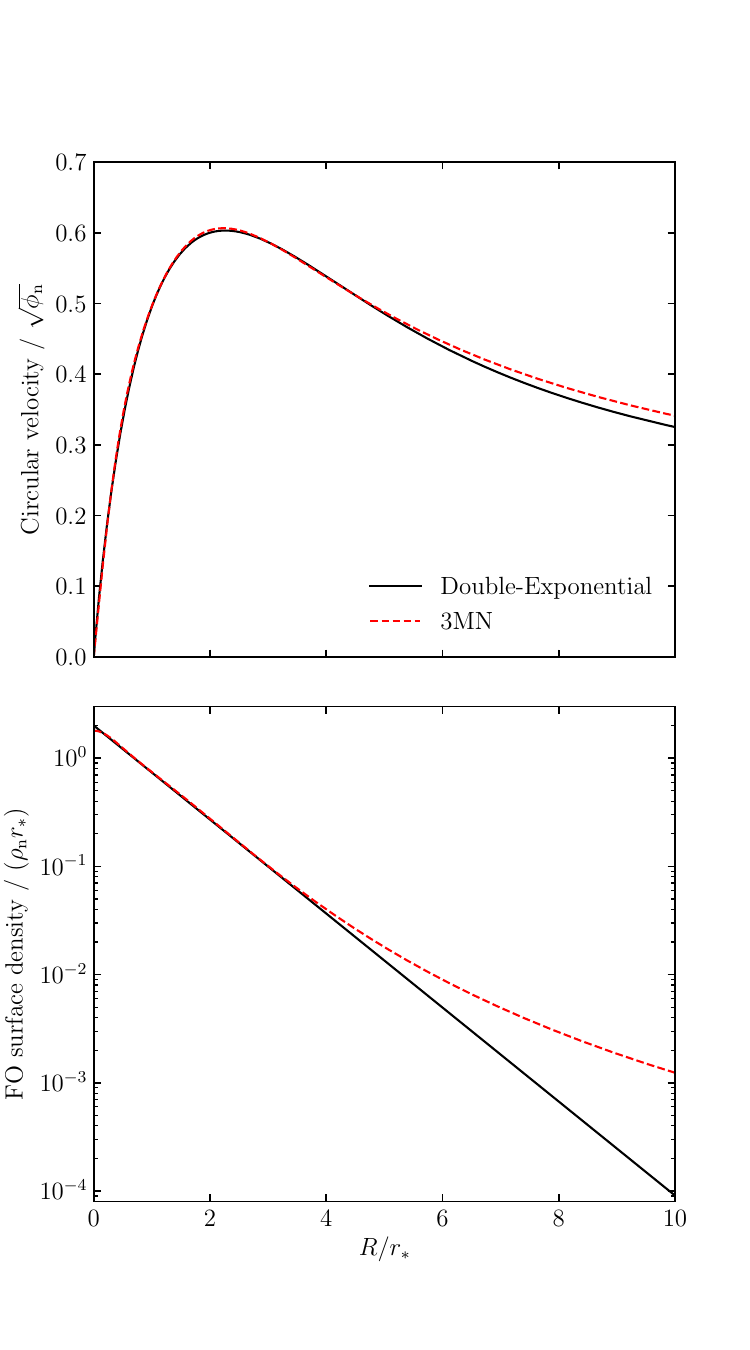}}
 \vspace{-.9cm}
 \caption{Circular velocity (top) and face-on surface density (bottom) profiles of the two models of Table \ref{tab:DExp_3MN_parameters}; these plots can be compared with Figures 3 and 7 in \citet{smith2015}.}
 \label{fig:DExp_3MN}
\end{figure}

\begin{figure*} % figure* to use both two columns 
 \centering 
 \subfloat{\includegraphics[width=\textwidth]{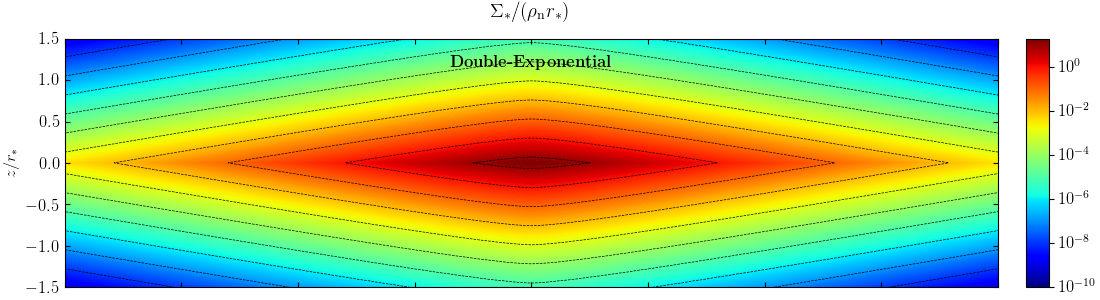}} \\ \vspace{-.2cm}
 \subfloat{\includegraphics[width=\textwidth]{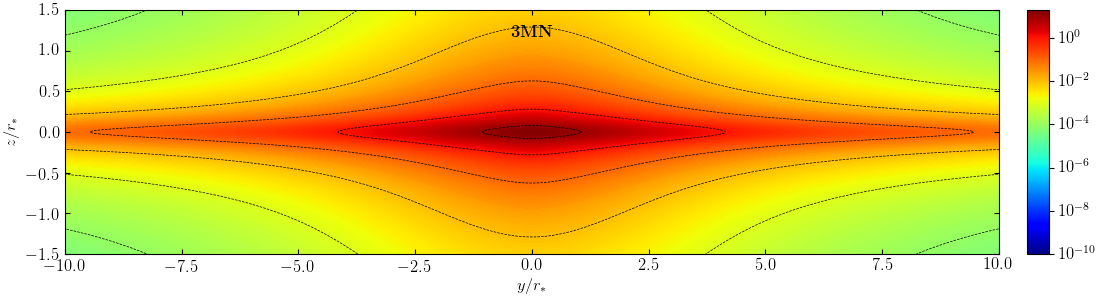}} \\
 \caption{Edge-on surface density distributions $\Sigmastar$ of the
   two models of Table \ref{tab:DExp_3MN_parameters}. The dotted
   contours are spaced by 1 dex.}
 \label{fig:DExp_3MN_surfdens}
\end{figure*}

We compute the potential for the double-exponential disc both with the
standard method in equation \eqref{eq:ellpot}, and with the much
faster integration of equation \eqref{eq:besselpot}, finding perfect agreement. As a safety check of the reconstructed 3MN model, we compare the circular velocity in the equatorial plane of the double-exponential disc and of its 3MN fit (Figure \ref{fig:DExp_3MN}, top panel), that can be compared with Figure 3 of \citet{smith2015}, and the FO surface density profiles of the two models (Figure \ref{fig:DExp_3MN}, bottom panel) that in turn can be compared with their Figure 7. The circular velocity of the exponential disc is almost perfectly reproduced over the explored radial range, while the reproduction of the FO surface density is less satisfactory, an unavoidable consequence of the everywhere positive decomposition adopted. For completeness, in Figure \ref{fig:DExp_3MN_surfdens}, we present the EO surface density distributions of the two models. As expected, the two distributions are quite different in the outer regions, especially for increasing distance from the equatorial plane, where the 3MN model produces higher surface density values. Consequently, also the kinematical fields obtained from the solution of the Jeans equations are expected to show significant differences, especially at high $|z|$.

\begin{figure*} % figure* to use both two columns 
 \centering 
 \subfloat{\includegraphics[width=\textwidth]{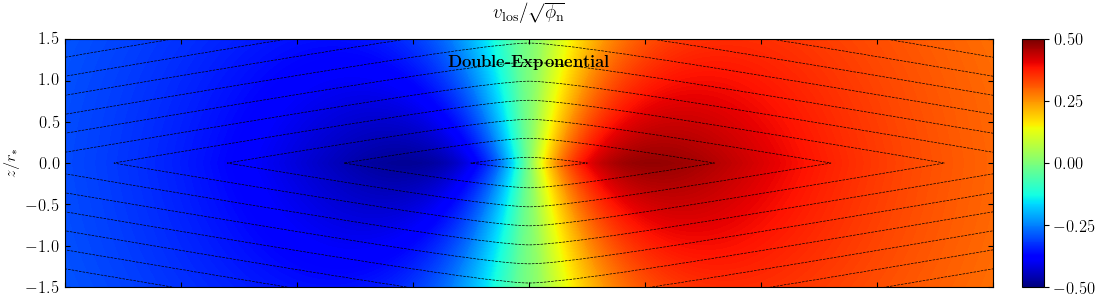}} \\ %\vspace{-.5cm}
 \subfloat{\includegraphics[width=\textwidth]{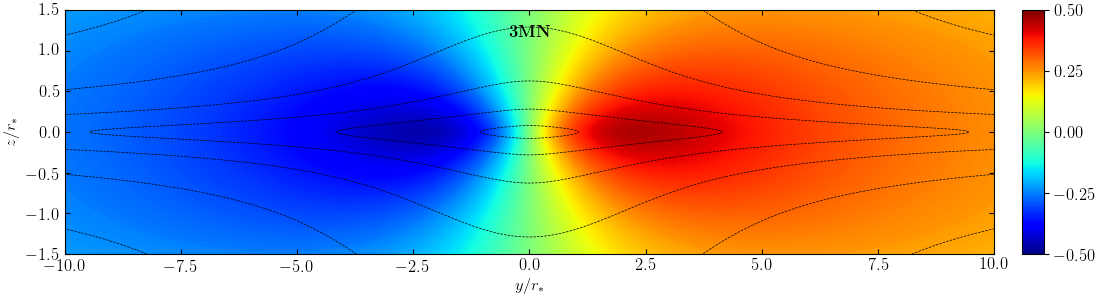}} \\
 \subfloat{\includegraphics[width=\textwidth]{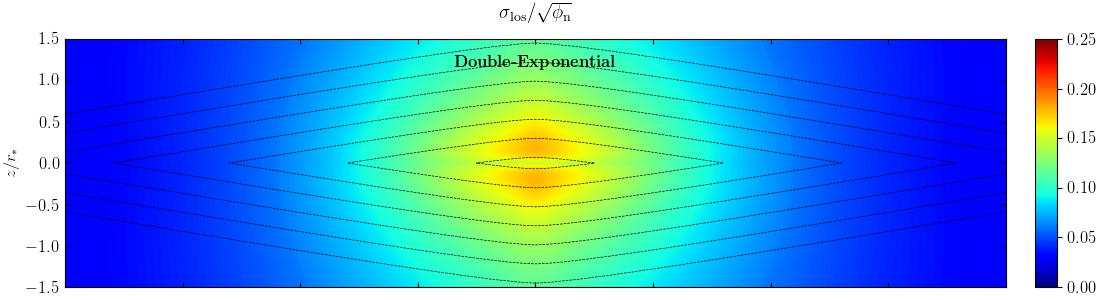}} \\ %\vspace{-.5cm}
 \subfloat{\includegraphics[width=\textwidth]{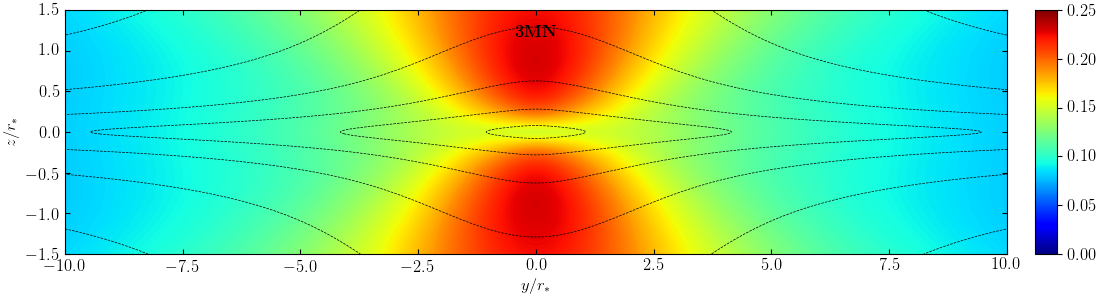}}
 \caption{Edge-on projected rotational velocity
   $\upsilon_\mathrm{los}$, and velocity
   dispersion $\sigma_\mathrm{los}$, normalised
   to ${\sqrt{\phin}}$, for the two models of Table
   \ref{tab:DExp_3MN_parameters}. The dotted contours are the same as
   in Figure \ref{fig:DExp_3MN_surfdens}. }
 \label{fig:DExp_3MN_losvel}
\end{figure*}

We use our procedure in JASMINE2 to evaluate these differences, a
problem left open by the studies of \citet{smith2015} and
\citet{rojas-nino2016}; we adopt for simplicity the case of the
isotropic rotator, without a DM halo and a central BH. In Figure
\ref{fig:DExp_3MN_losvel}, the EO projected rotational velocity and
velocity dispersion are shown. The fields $\upsilon_\mathrm{los}$ of
the two models look remarkably similar, also outside the equatorial
plane. In particular, the percent error of the 3MN model with respect
to the double-exponential model, in the equatorial plane, is $<9\%$
out to $4\,\Rd$, and $<14\%$ out to $10\,\Rd$. This quite satisfactory
result is not obvious a priori, since $\vphi$, at variance with
$\upsilon_{\rm c}$, is not a function of the potential only, but it
also depends on the velocity dispersion via the asymmetric
drift. Therefore, the excellent agreement of $\upsilon_{\rm c}$ in
Figure \ref{fig:DExp_3MN} is not a guarantee that also $\vphi$, and
its projection $\upsilon_\mathrm{los}$, are well reproduced by the 3MN
density fit. The reproduction of $\upsilon_\mathrm{los}$ outside the
equatorial plane is still quite good, with a slightly higher
discrepancy at increasing $|z|$, as expected, but improving for larger
galactocentric distances; for example, at $z=\Rd$, the percent error
is $<23\%$ out to $4\,\Rd$, reducing to $<18\%$ out to $10\,\Rd$. The
situation is different for $\sigma_\mathrm{los}$: the two fields are
significantly different, even in the equatorial plane, with
the 3MN model showing values up to a factor of 2 larger than those of
the double-exponential model. Moreover, the velocity dispersion of the
3MN model near the rotation axis presents a characteristic
hourglass-shaped distribution \citep[see also][]{negri14}, only barely
detectable at the very centre for the double-exponential model. Notice
that this feature is not observed in the maps of Figure
\ref{fig:losvel} for the JHD and JLD models, even if they also contain
a MN component, due to the addition of a stellar halo and a DM halo,
and to the different kinematical decompositions adopted \citep[see
also the discussion in][]{smet_pos15}. These experiments suggest
caution when adopting the 3MN representation
to interpret the observed velocity dispersion of disc galaxies.

%%%%%%%%%%%%%%%%%%%%%%%%%%%%%%%%%%%%%%%%%%%%%%%%%%%%%%%
\begin{table}
 \centering 
 \begin{tabular}{cc}
  \toprule\toprule 
  Model &  \\
  \midrule\midrule 
  Double-Exponential disc & $\alpha=\Rd/\rstar=1$ \\
  & $\beta=h/\rstar=0.1$ \\
  & $\RMd=1$ \\
  & $k=1$ \\
  \midrule
  3MN fit \citep{smith2015} & $\tilde b=0.12$ \\
  & $q_1=4.64, \quad q_2=21.42, \quad q_3=18.67$ \\
  & $\RM_1=0.16, \quad \RM_2=-5.77, \quad \RM_3=6.72$ \\
  & $k=1$ \\
  \bottomrule\bottomrule 
 \end{tabular}
 \caption{The parameters of the double-exponential disc and its 3MN fit from \citet{smith2015}. For the meaning of the parameters of the double-exponential disc, see Section \ref{sec:results_DExp_3MN} and Appendix \ref{app:expdisc}. For the 3MN model, we adopt the same notation of Section \ref{sec:JHD&JLD}, with the same $\tilde b$ for all the three components, $q_i=\tilde a_i/\tilde b$, and $\RM_i=\Mstari/\Mstar$. The Jeans equations are solved in the isotropic case, with constant Satoh parameter $k=1$.}
 \label{tab:DExp_3MN_parameters}
\end{table}
%%%%%%%%%%%%%%%%%%%%%%%%%%%%%%%%%%%%%%%%%%%%%%%%%%%%%%%

%%%%%%%%%%%%%%%%%%%%%%%%%%%%%%%%%%%%%%%%%%%%%%%%%%%%%%%%%%%%%%%%%%%%%
%%%%%%%%%%%%%%%%%%%%%%%%%%%%%%%%%%%%%%%%%%%%%%%%%%%%%%%%%%%%%%%%%%%%%

%%%%%%%%%%%%%%%%%%%%%%%%%%%%%%%%%%%%%%%%%%%%%%%%%%%%%%%%%%%%%%%%%%%%%
%%%%%%%%%%%%%%%%%%%%%%%%%%%%%%%%%%%%%%%%%%%%%%%%%%%%%%%%%%%%%%%%%%%%%

%%%%%%%%%%%%%%%%%%%%%%%%%%%%%%%%%%%%%%%%%%%%%%%%%%%%%%%%%%%%%%%%%%%%%
%%%%%%%%%%%%%%%%%%%%%%%%%%%%%%%%%%%%%%%%%%%%%%%%%%%%%%%%%%%%%%%%%%%%%

\section{Conclusions} 
\label{sec:concl}

We presented the theoretical framework for an efficient Jeans modeling
of multi-component axisymmetric galaxies, and its numerical
implementation in the code JASMINE2, significantly upgraded from
  its original version JASMINE \citep{pos13}. In this framework the
models can include an arbitrary number of stellar components, with
different structural, dynamical and stellar population properties, a DM halo, and a central BH. The structural and dynamical properties of each stellar component can be mass- or luminosity-weighted, and projected on the plane of the
sky. The internal dynamics of each stellar component is implicitly
described by a 2-integral DF (in general different for each component), so that a phenomenological decomposition of the azimuthal velocity field must be chosen. For each component, we can adopt the \citet{satoh80} $k$-decomposition, where $\vphii = k_i \sqrt{\Di}$, or a generalised $k$-decomposition, where $\vphii = k_i \sqrt{\Di+\sigma_i^2}$; furthermore, the parameter $k_i$ can be constant or position-dependent. The generalised decomposition allows for the modeling of systems with $\Delta_i<0$ (as may happen for density distributions elongated along the symmetry axis). The presented scheme can be easily extended to more general velocity decompositions, such as that introduced by \citet{capp08}, where the underlying DF depends on 3 integrals of motion.

In the numerical implementation, the gravitational potential of the density components is computed by default in terms of complete elliptic integrals (equation \ref{eq:ellpot}), a very accurate but quite time-expensive approach, that can become impractical especially when dealing with the exploration of the parameter space of multi-component models. To reduce the computational time, we fully exploited the scalings allowed by the Poisson and the Jeans equations, and by the projection formulae. The resulting scheme led to an organisation of JASMINE2 in two logically distinct parts: the Potential and Jeans Solver and the Post-Processing (PP). In practice, once the structural properties of the scaled stellar and DM distributions are assigned, the code computes, with a single run of the Potential and Jeans Solver, the scaled solutions of the Jeans equations, defining a \textit{family} of models. The scaled Jeans solutions are then combined in PP, with the desired mass and luminosity weights, and the choice of appropriate kinematical
decompositions, and then projected. The PP procedure can be performed
several times, obtaining different specific models in the same
family. Finally, for each model, the two physical scales $\Mstar$ and
$\rstar$ can be assigned. A further benefit of the presented approach
is the possibility to gain a full understanding of the
role of each density component in determining the resulting
kinematical fields of the galaxy. For special density distributions, a further reduction of computational time is obtained by evaluating the potential with specific integral formulae, such as the Chandrasekhar formula for ellipsoidal distributions, and integrals involving Bessel functions for factorised disc distributions (see Appendix \ref{app:expdisc}).

In order to illustrate the features of our modeling procedure, we presented three galaxy models, composed of two stellar components, a spherical NFW DM halo and a central supermassive BH. In the JJE model, the total spherical stellar profile and one ellipsoidal stellar component, both Jaffe models, are assigned; the second stellar component is given by
their difference. This model, when the DM halo is absent, has several
properties available in analytical form (in particular in \citetalias{cmpz21}), and thus has been used to test the procedure and the code. The JHD model consists of a large and massive MN stellar disc, coupled with an ellipsoidal Jaffe stellar model, that can be seen as the stellar halo of a disc galaxy. In the JLD model, an ellipsoidal Jaffe component dominates in mass, and a MN stellar disc is small, inner and counter-rotating, as sometimes found in early-type galaxies.

As a fourth application, we explored the accuracy of one of the
  3MN decompositions proposed by \citet{smith2015} to reproduce the
  kinematical fields of double-exponential discs. We confirmed the
  excellent agreement of the rotation curves of the two models in the
  equatorial plane, at least out to $\sim 10R_d$. We also found that
  $\upsilon_\mathrm{los}$ tends to be larger for the
  double-exponential disc than for its 3MN representation, but overall
  the agreement is rather good, even outside the equatorial plane. A
  different situation is found for $\sigma_\mathrm{los}$: the values
  are significantly larger in the 3MN model, which also presents a
  characteristic hourglass-shaped vertical distribution. Some care is
  thus recommended when using a 3MN decomposition to infer the properties of observed disc galaxies.

Ongoing applications of the presented modeling procedure, and in particular of JASMINE2, include the building of
multi-component galaxy models for numerical simulations of gas flows
in galaxies (e.g. \citealt{negri_pos_pel_ciot14, gan_ciot19, gan_choi19}; \citetalias{ciot_gan21} et al. in preparation); the study
of circumnuclear stellar discs (also with counter-rotation, see
e.g. \citealt{morelli04, kraj15, mitzkus17, sormani20}; see also
\citealt{capp-rev16}); a systematic exploration of galaxy models
constrained to lie on the major observed Scaling Laws, extending the
statistical approach pioneered in \citet{bertin-ciot-delprinc02} and
\citet{lanz-ciot03}.
 
\section*{Acknowledgements}

We are grateful to Antonio Mancino for independent checks of the results
of JJE models. We thank the anonymous referee for useful comments that improved the paper.

\section*{Data availability}

The data underlying this article were produced by the
authors. They will be shared under reasonable request to the corresponding author.

\bibliographystyle{mnras} % mnras.bst
\bibliography{ms} % ms.bib

\begin{thebibliography}{}
\makeatletter
\relax
\def\mn@urlcharsother{\let\do\@makeother \do\$\do\&\do\#\do\^\do\_\do\%\do\~}
\def\mn@doi{\begingroup\mn@urlcharsother \@ifnextchar [ {\mn@doi@}
  {\mn@doi@[]}}
\def\mn@doi@[#1]#2{\def\@tempa{#1}\ifx\@tempa\@empty \href
  {http://dx.doi.org/#2} {doi:#2}\else \href {http://dx.doi.org/#2} {#1}\fi
  \endgroup}
\def\mn@eprint#1#2{\mn@eprint@#1:#2::\@nil}
\def\mn@eprint@arXiv#1{\href {http://arxiv.org/abs/#1} {{\tt arXiv:#1}}}
\def\mn@eprint@dblp#1{\href {http://dblp.uni-trier.de/rec/bibtex/#1.xml}
  {dblp:#1}}
\def\mn@eprint@#1:#2:#3:#4\@nil{\def\@tempa {#1}\def\@tempb {#2}\def\@tempc
  {#3}\ifx \@tempc \@empty \let \@tempc \@tempb \let \@tempb \@tempa \fi \ifx
  \@tempb \@empty \def\@tempb {arXiv}\fi \@ifundefined
  {mn@eprint@\@tempb}{\@tempb:\@tempc}{\expandafter \expandafter \csname
  mn@eprint@\@tempb\endcsname \expandafter{\@tempc}}}

\bibitem[\protect\citeauthoryear{{Bertin}}{{Bertin}}{2014}]{bertin14}
{Bertin} G.,  2014, {Dynamics of Galaxies}.
Cambridge University Press, \mn@doi{10.1017/CBO9780511731990}

\bibitem[\protect\citeauthoryear{{Bertin}, {Ciotti}  \& {Del
  Principe}}{{Bertin} et~al.}{2002}]{bertin-ciot-delprinc02}
{Bertin} G.,  {Ciotti} L.,   {Del Principe} M.,  2002, \mn@doi [\aap]
  {10.1051/0004-6361:20020248}, \href
  {https://ui.adsabs.harvard.edu/abs/2002A&A...386..149B} {386, 149}

\bibitem[\protect\citeauthoryear{{Binney} \& {Tremaine}}{{Binney} \&
  {Tremaine}}{2008}]{b&t08}
{Binney} J.,  {Tremaine} S.,  2008, {Galactic Dynamics: Second Edition}.
Princeton University Press

\bibitem[\protect\citeauthoryear{{Cappellari}}{{Cappellari}}{2008}]{capp08}
{Cappellari} M.,  2008, \mn@doi [MNRAS] {10.1111/j.1365-2966.2008.13754.x},
  \href {http://adsabs.harvard.edu/abs/2008MNRAS.390...71C} {390, 71}

\bibitem[\protect\citeauthoryear{{Cappellari}}{{Cappellari}}{2016}]{capp-rev16}
{Cappellari} M.,  2016, \mn@doi [ARAA] {10.1146/annurev-astro-082214-122432},
  \href {http://adsabs.harvard.edu/abs/2016ARA%26A..54..597C} {54, 597}

\bibitem[\protect\citeauthoryear{{Caravita}}{{Caravita}}{2022}]{caravita_phd}
{Caravita} C.,  2022, {PhD Thesis}.
Bologna University

\bibitem[\protect\citeauthoryear{{Ciotti}}{{Ciotti}}{2021}]{ciotti21}
{Ciotti} L.,  2021, {Introduction to Stellar Dynamics}.
Cambridge University Press

\bibitem[\protect\citeauthoryear{{Ciotti} \& {Bertin}}{{Ciotti} \&
  {Bertin}}{2005}]{ciot-bertin05}
{Ciotti} L.,  {Bertin} G.,  2005, \mn@doi [\aap] {10.1051/0004-6361:20042123},
  \href {https://ui.adsabs.harvard.edu/abs/2005A&A...437..419C} {437, 419}

\bibitem[\protect\citeauthoryear{{Ciotti} \& {Pellegrini}}{{Ciotti} \&
  {Pellegrini}}{1996}]{ciot-pel96}
{Ciotti} L.,  {Pellegrini} S.,  1996, \mn@doi [MNRAS]
  {10.1093/mnras/279.1.240}, \href
  {http://adsabs.harvard.edu/abs/1996MNRAS.279..240C} {279, 240}

\bibitem[\protect\citeauthoryear{{Ciotti}, {Mancino}, {Pellegrini}  \& {Ziaee
  Lorzad}}{{Ciotti} et~al.}{2021}]{cmpz21}
{Ciotti} L.,  {Mancino} A.,  {Pellegrini} S.,   {Ziaee Lorzad} A.,  2021,
  \mn@doi [\mnras] {10.1093/mnras/staa3338}, \href
  {https://ui.adsabs.harvard.edu/abs/2021MNRAS.500.1054C} {500, 1054}

\bibitem[\protect\citeauthoryear{{Ciotti}, {Gan}, {Ostriker}, {Pellegrini},
  {Caravita}  \& {Mancino}}{{Ciotti} et~al.}{2022}]{ciot_gan21}
{Ciotti} L.,  {Gan} Z.,  {Ostriker} J.~P.,  {Pellegrini} S.,  {Caravita} C.,
  {Mancino} A.,  2022, in preparation

\bibitem[\protect\citeauthoryear{{Flynn}, {Sommer-Larsen}  \&
  {Christensen}}{{Flynn} et~al.}{1996}]{flynn96}
{Flynn} C.,  {Sommer-Larsen} J.,   {Christensen} P.~R.,  1996, \mn@doi [\mnras]
  {10.1093/mnras/281.3.1027}, \href
  {https://ui.adsabs.harvard.edu/abs/1996MNRAS.281.1027F} {281, 1027}

\bibitem[\protect\citeauthoryear{{Gan}, {Ciotti}, {Ostriker}  \& {Yuan}}{{Gan}
  et~al.}{2019a}]{gan_ciot19}
{Gan} Z.,  {Ciotti} L.,  {Ostriker} J.~P.,   {Yuan} F.,  2019a, \mn@doi [\apj]
  {10.3847/1538-4357/ab0206}, \href
  {https://ui.adsabs.harvard.edu/abs/2019ApJ...872..167G} {872, 167}

\bibitem[\protect\citeauthoryear{{Gan}, {Choi}, {Ostriker}, {Ciotti}  \&
  {Pellegrini}}{{Gan} et~al.}{2019b}]{gan_choi19}
{Gan} Z.,  {Choi} E.,  {Ostriker} J.~P.,  {Ciotti} L.,   {Pellegrini} S.,
  2019b, \mn@doi [\apj] {10.3847/1538-4357/ab1007}, \href
  {https://ui.adsabs.harvard.edu/abs/2019ApJ...875..109G} {875, 109}

\bibitem[\protect\citeauthoryear{{Jaffe}}{{Jaffe}}{1983}]{jaffe83}
{Jaffe} W.,  1983, \mn@doi [MNRAS] {10.1093/mnras/202.4.995}, \href
  {http://adsabs.harvard.edu/abs/1983MNRAS.202..995J} {202, 995}

\bibitem[\protect\citeauthoryear{{Kormendy} \& {Ho}}{{Kormendy} \&
  {Ho}}{2013}]{korm-ho13}
{Kormendy} J.,  {Ho} L.~C.,  2013, \mn@doi [\araa]
  {10.1146/annurev-astro-082708-101811}, \href
  {https://ui.adsabs.harvard.edu/abs/2013ARA&A..51..511K} {51, 511}

\bibitem[\protect\citeauthoryear{{Krajnovi{\'c}} et~al.,}{{Krajnovi{\'c}}
  et~al.}{2015}]{kraj15}
{Krajnovi{\'c}} D.,  et~al., 2015, \mn@doi [\mnras] {10.1093/mnras/stv958},
  \href {https://ui.adsabs.harvard.edu/abs/2015MNRAS.452....2K} {452, 2}

\bibitem[\protect\citeauthoryear{{Kuzmin}}{{Kuzmin}}{1956}]{kuzmin56}
{Kuzmin} G.~G.,  1956, \azh, 33, 27

\bibitem[\protect\citeauthoryear{{Lanzoni} \& {Ciotti}}{{Lanzoni} \&
  {Ciotti}}{2003}]{lanz-ciot03}
{Lanzoni} B.,  {Ciotti} L.,  2003, \mn@doi [\aap] {10.1051/0004-6361:20030488},
  \href {https://ui.adsabs.harvard.edu/abs/2003A&A...404..819L} {404, 819}

\bibitem[\protect\citeauthoryear{{Maraston}}{{Maraston}}{2005}]{mar05}
{Maraston} C.,  2005, \mn@doi [MNRAS] {10.1111/j.1365-2966.2005.09270.x}, \href
  {http://adsabs.harvard.edu/abs/2005MNRAS.362..799M} {362, 799}

\bibitem[\protect\citeauthoryear{{Mitzkus}, {Cappellari}  \&
  {Walcher}}{{Mitzkus} et~al.}{2017}]{mitzkus17}
{Mitzkus} M.,  {Cappellari} M.,   {Walcher} C.~J.,  2017, \mn@doi [\mnras]
  {10.1093/mnras/stw2677}, \href
  {https://ui.adsabs.harvard.edu/abs/2017MNRAS.464.4789M} {464, 4789}

\bibitem[\protect\citeauthoryear{{Miyamoto} \& {Nagai}}{{Miyamoto} \&
  {Nagai}}{1975}]{MN}
{Miyamoto} M.,  {Nagai} R.,  1975, \pasj, \href
  {https://ui.adsabs.harvard.edu/abs/1975PASJ...27..533M} {27, 533}

\bibitem[\protect\citeauthoryear{{Morelli} et~al.,}{{Morelli}
  et~al.}{2004}]{morelli04}
{Morelli} L.,  et~al., 2004, \mn@doi [\mnras]
  {10.1111/j.1365-2966.2004.08236.x}, \href
  {http://adsabs.harvard.edu/abs/2004MNRAS.354..753M} {354, 753}

\bibitem[\protect\citeauthoryear{{Navarro}, {Frenk}  \& {White}}{{Navarro}
  et~al.}{1996}]{NFW}
{Navarro} J.~F.,  {Frenk} C.~S.,   {White} S.~D.~M.,  1996, \mn@doi [ApJ]
  {10.1086/177173}, \href {http://adsabs.harvard.edu/abs/1996ApJ...462..563N}
  {462, 563}

\bibitem[\protect\citeauthoryear{{Negri}, {Ciotti}  \& {Pellegrini}}{{Negri}
  et~al.}{2014a}]{negri14}
{Negri} A.,  {Ciotti} L.,   {Pellegrini} S.,  2014a, \mn@doi [\mnras]
  {10.1093/mnras/stt2505}, \href
  {https://ui.adsabs.harvard.edu/abs/2014MNRAS.439..823N} {439, 823}

\bibitem[\protect\citeauthoryear{{Negri}, {Posacki}, {Pellegrini}  \&
  {Ciotti}}{{Negri} et~al.}{2014b}]{negri_pos_pel_ciot14}
{Negri} A.,  {Posacki} S.,  {Pellegrini} S.,   {Ciotti} L.,  2014b, \mn@doi
  [\mnras] {10.1093/mnras/stu1834}, \href
  {https://ui.adsabs.harvard.edu/abs/2014MNRAS.445.1351N} {445, 1351}

\bibitem[\protect\citeauthoryear{{Plummer}}{{Plummer}}{1911}]{plummer11}
{Plummer} H.~C.,  1911, \mn@doi [\mnras] {10.1093/mnras/71.5.460}, \href
  {https://ui.adsabs.harvard.edu/abs/1911MNRAS..71..460P} {71, 460}

\bibitem[\protect\citeauthoryear{{Posacki}, {Pellegrini}  \&
  {Ciotti}}{{Posacki} et~al.}{2013}]{pos13}
{Posacki} S.,  {Pellegrini} S.,   {Ciotti} L.,  2013, \mn@doi [MNRAS]
  {10.1093/mnras/stt898}, \href
  {http://adsabs.harvard.edu/abs/2013MNRAS.433.2259P} {433, 2259}

\bibitem[\protect\citeauthoryear{{Renzini} \& {Buzzoni}}{{Renzini} \&
  {Buzzoni}}{1986}]{renz-buz86}
{Renzini} A.,  {Buzzoni} A.,  1986, in {Chiosi} C.,  {Renzini} A.,  eds,
  Astrophysics and Space Science Library Vol. 122, Spectral Evolution of
  Galaxies. pp 195--231, \mn@doi{10.1007/978-94-009-4598-2_19}

\bibitem[\protect\citeauthoryear{{Rojas-Ni{\~n}o}, {Read}, {Aguilar}  \&
  {Delorme}}{{Rojas-Ni{\~n}o} et~al.}{2016}]{rojas-nino2016}
{Rojas-Ni{\~n}o} A.,  {Read} J.~I.,  {Aguilar} L.,   {Delorme} M.,  2016,
  \mn@doi [\mnras] {10.1093/mnras/stw846}, \href
  {https://ui.adsabs.harvard.edu/abs/2016MNRAS.459.3349R} {459, 3349}

\bibitem[\protect\citeauthoryear{{Satoh}}{{Satoh}}{1980}]{satoh80}
{Satoh} C.,  1980, PASJ, \href
  {http://adsabs.harvard.edu/abs/1980PASJ...32...41S} {32, 41}

\bibitem[\protect\citeauthoryear{{Smet}, {Posacki}  \& {Ciotti}}{{Smet}
  et~al.}{2015}]{smet_pos15}
{Smet} C.~O.,  {Posacki} S.,   {Ciotti} L.,  2015, \mn@doi [\mnras]
  {10.1093/mnras/stv202}, \href
  {https://ui.adsabs.harvard.edu/abs/2015MNRAS.448.2921S} {448, 2921}

\bibitem[\protect\citeauthoryear{{Smith}, {Flynn}, {Candlish}, {Fellhauer}  \&
  {Gibson}}{{Smith} et~al.}{2015}]{smith2015}
{Smith} R.,  {Flynn} C.,  {Candlish} G.~N.,  {Fellhauer} M.,   {Gibson} B.~K.,
  2015, \mn@doi [\mnras] {10.1093/mnras/stv228}, \href
  {https://ui.adsabs.harvard.edu/abs/2015MNRAS.448.2934S} {448, 2934}

\bibitem[\protect\citeauthoryear{{Sormani}, {Magorrian}, {Nogueras-Lara},
  {Neumayer}, {Sch{\"o}nrich}, {Klessen}  \& {Mastrobuono-Battisti}}{{Sormani}
  et~al.}{2020}]{sormani20}
{Sormani} M.~C.,  {Magorrian} J.,  {Nogueras-Lara} F.,  {Neumayer} N.,
  {Sch{\"o}nrich} R.,  {Klessen} R.~S.,   {Mastrobuono-Battisti} A.,  2020,
  \mn@doi [\mnras] {10.1093/mnras/staa2785}, \href
  {https://ui.adsabs.harvard.edu/abs/2020MNRAS.499....7S} {499, 7}

\bibitem[\protect\citeauthoryear{{Yoon}, {Yuan}, {Ostriker}, {Ciotti}  \&
  {Zhu}}{{Yoon} et~al.}{2019}]{Yoon19}
{Yoon} D.,  {Yuan} F.,  {Ostriker} J.~P.,  {Ciotti} L.,   {Zhu} B.,  2019,
  \mn@doi [\apj] {10.3847/1538-4357/ab45e8}, \href
  {https://ui.adsabs.harvard.edu/abs/2019ApJ...885...16Y} {885, 16}

\makeatother
\end{thebibliography}

\appendix

\section{Positivity condition for JJE models}
\label{app:pos_cond}

The stellar component $\rhos$, for the 2-component ellipsoidal models
described in Section \ref{sec:JJE}, is given by the difference of an
assigned total $\rhostar$ and an assigned $\rhof$. This approach
naturally leads to discuss the positivity of $\rhos$, with a treatment
similar to that followed in the appendix of \citetalias{cmpz21}, and
references therein. We generalise equation \eqref{eq:rho2_diff} as
\begin{equation} 
{\rhos(R,z)\over (3-\gamma)\rhon} = {\xi\over q m^\gamma (\xi+m)^{4-\gamma}}
 - {\RMo \xi_1\over q_1 m_1^\gamma(\xi_1+m_1)^{4-\gamma}},
 \label{eq:non-necond_first}
\end{equation}
recovering the case of JJE models for $\gamma=2$. In order to discuss
the positivity condition for $\rhos$, we use spherical coordinates, so
that $(R,z)=r(\sin\theta, \cos\theta)$ and
\begin{equation}
 m = s \Omega, \quad m_1 = s \Omega_1, \quad s \equiv {r\over\rstar},
\end{equation}
where
\begin{equation} 
 \Omega^2 \equiv \sin^2\theta+{\cos^2\theta\over q^2}, \quad
 \Omega_1^2 \equiv \sin^2\theta+{\cos^2\theta\over q_1^2}.
 \label{eq:def_s_omega}
\end{equation}
The positivity of $\rhos$ reduces to a condition on $\RMo$, given by
\begin{equation} 
  \RMo \leq \RMax \equiv \inf_\mathcal{I}\left[
    {\xi q_1\over\xi_1 q} \left( {\Omega_1\over\Omega} \right)^{\gamma}
   \left({\xi_1+s\Omega_1\over \xi+s\Omega}\right)^{4-\gamma}
\right],
 \label{eq:pos_cond_second}
\end{equation}
over the rectangular region
$\mathcal{I} \equiv \{s\geq 0, 0\leq\theta\leq\pi/2\}$ in the
$(s,\theta)$ plane. Following the discussion in \citetalias{cmpz21},
we determine
\begin{equation}
 \RMax = \min\left(\RM_{\rm c}, \RM_{\infty}, \RM_0, \RM_{\pi/2},
   \RM_{\rm int}\right),
 \label{eq:pos_fin}
\end{equation}
where the first four quantities refer to the minimum value of the
r.h.s. of equation \eqref{eq:pos_cond_second} over the boundaries of
$\mathcal{I}$, and $\RM_{\rm int}$ is the value of a minimum (if it
exists) in the interior of $\mathcal{I}$. When $q_1\neq q$, it is
simple to show that no critical points can exist in the interior of
$\mathcal{I}$, and so the discussion reduces to the boundaries of
$\mathcal{I}$: geometrically, $\RMax$ can be reached only at the
centre ($s=0$, $\RM_{\rm c}$), at infinity ($s\to\infty$,
$\RM_{\infty}$), along the symmetry axis ($\theta=0$, $\RM_0$), or on
the equatorial plane ($\theta=\pi/2$, $\RM_{\pi/2}$).

We begin with $\RM_{\rm c}$ and $\RM_{\infty}$, obtaining
\begin{equation} 
 \RM_{\rm c} = {\xi_1^{3-\gamma}q_1\over\xi^{3-\gamma}q}
     \min_{0\leq\theta\leq\pi/2}\left({\Omega_1\over\Omega}\right)^\gamma,
 \label{eq:rc}
\end{equation}
\begin{equation} 
 \RM_{\infty} = {\xi q_1\over\xi_1 q}
     \min_{0\leq\theta\leq\pi/2}
     \left({\Omega_1\over\Omega}\right)^4.
 \label{eq:rinf}
\end{equation}
Now, from equation \eqref{eq:def_s_omega}, it is easy to show that for
a generic $\alpha\geq 0$, the function $(\Omega_1/\Omega)^\alpha$
reaches its minimum at $\theta=\pi/2$ if $q_1\leq q$, and at
$\theta=0$ if $q\leq q_1$, so that
\begin{equation}
  \min_{0\leq\theta\leq\pi/2}\left({\Omega_1\over\Omega}\right)^\alpha = 
 \begin{cases}
   1,\quad q_1\leq q, \\ \\
   \displaystyle{
  \left({q\over q_1}\right)^\alpha,\quad q\leq q_1,}
 \end{cases}
\end{equation} 
and the conditions in equations \eqref{eq:rc} and \eqref{eq:rinf} can be finally
summarised as
\begin{equation}
\RM_{\rm c} = {\xi_1^{3-\gamma}q_1\over\xi^{3-\gamma}q}
\min\left(1,{q^{\gamma}\over q_1^{\gamma}}\right),
\end{equation}
\begin{equation}
\RM_{\infty} = {\xi q_1\over\xi_1 q} \min\left(1, {q^4\over q_1^4} \right).
\end{equation}
Along the symmetry axis, and in the equatorial plane, condition
\eqref{eq:pos_cond_second} becomes
\begin{equation}
 \RM_0 = {\xi q^3\over \xi_1 q_1^3}\inf_{0\leq s<\infty}
    \left( {\xi_1 q_1+s\over \xi q +s} \right)^{4-\gamma},
\end{equation}
\begin{equation}
 \RM_{\pi/2} = {\xi q_1\over \xi_1 q}\inf_{0\leq s<\infty}\left({\xi_1+s\over\xi+s}\right)^{4-\gamma},
\end{equation}
and simple algebra finally shows that the results can be summarised as 
\begin{equation}
  \RM_0 = {\xi q^3\over \xi_1 q_1^3}\min\left[1,\left({\xi_1 q_1\over
  \xi q}\right)^{4-\gamma}\right],
\end{equation}
\begin{equation}
 \RM_{\pi/2}= {\xi q_1\over \xi_1 q} \min\left(1,{\xi_1^{4-\gamma}\over\xi^{4-\gamma}}\right).
\end{equation}

For the JJE models in Section \ref{sec:JJE}, with $\xi_1<\xi$, $q_1<q$, and $\gamma=2$, the positivity condition \eqref{eq:pos_fin} becomes
\begin{equation}
 \RMo \leq \RMax = {\xi_1 q_1\over \xi q}.
 \label{eq:pos_cond_final}
\end{equation}

\section{Potential of factorised exponential discs}
\label{app:expdisc}

Due to the importance in applications, here we summarise the main
results about the numerical evaluation of the potential produced by factorised exponential discs as in equation \eqref{eq:fact_exp_rho}, by using the technique of Bessel functions. From equations (2.103) and (2.114) in \citet{ciotti21}, the potential can be easily written in full generality as 
\begin{equation}
  \phi(R,z) =-2\pi 
  G\rho_0\rstar^2\alpha^2\beta\int_0^{\infty}{\Jzer(\lambda\tildR)\hat 
    V(\lambda\beta,\tildz/\beta)\over (1+\alpha^2\lambda^2)^{3/2}}\dlambda,
  \label{eq:besselpot}
\end{equation}
where $\Jzer$ is a Bessel function of the first kind,
$\alpha=\Rd/\rstar$, $\beta=h/\rstar$, $\tildR=R/\rstar$,
$\tildz=z/\rstar$, and finally
\begin{equation}
\hat V(\gamma, x)\equiv \int_{-\infty}^{\infty}{\rm e}^{-\gamma\vert
  \vert x\vert-t\vert}V(\vert t\vert)\dt.
\end{equation}
Therefore, once the function $\hat V$ is known
analytically, the integration for each grid point $(\tildR,\tildz)$
reduces to a fast 1-dimensional integration, instead of the more time-consuming 2-dimensional integration that would be required when using the standard formula based on complete elliptic integrals (equation \ref{eq:ellpot}), or the alternative equation based on modified Bessel functions (equation 2.170 in \citetalias{b&t08}; equation 4 in \citealt{smith2015}).

For the three discs considered in \citet{smith2015}, we define $\RMd=M_{\rm d}/\Mstar$, where $M_{\rm d}$ is the total mass of the disc, so that for the double-exponential disc considered in Section \ref{sec:results_DExp_3MN}, we have
\begin{equation}
  \rhostar(R,z)={\rhon\RMd\over\alpha^2\beta}{\rm e}^{-\tildR/\alpha-\vert\tildz\vert/\beta},
\quad\hat V(\gamma,x)=
  {2(\gamma{\rm e}^{-\vert x\vert}-{\rm e}^{-\gamma \vert x\vert})\over\gamma^2 -1},
  \label{eq:DExp_rho}
\end{equation}
for $\gamma\neq 1$,  and $\hat V(1,x)=(1+\vert x\vert){\rm e}^{-\vert x\vert}$. For completness, we also report the formulae for the razor-thin exponential disc,
\begin{equation}
  \rhostar(R,z)={2\rhon\RMd\over\alpha^2\beta}
    {\rm  e}^{-\tildR/\alpha}\delta(\tildz/\beta),
\qquad \hat V(\gamma,x)={\rm e}^{-\gamma\vert x\vert},
  \label{eq:razor-thin_rho}
\end{equation}
where $\delta$ is the Dirac-$\delta$ function, and for the ''pseudo-isothermal'' exponential disc,
\begin{equation}
  \rhostar(R,z) = {2^{2-a} \rhon\RMd\over\alpha^2\beta\,{\rm 
      B}(a/2,a/2)} 
    {{\rm  e}^{-\tildR/\alpha}\over\cosh(\tildz/\beta)^a},  \quad a > 0,
\end{equation}
where ${\rm B}(x,y)$ is the Euler complete Beta function, and
\begin{equation}
  \begin{split}
    \hat V(\gamma,x) =
    & 2^{a-1}{\rm e}^{-\gamma\vert x\vert}{\rm B}\left({a+\gamma\over 2}, {a-\gamma\over 2}, {{\rm e}^{2\vert x\vert}\over 1+{\rm e}^{2\vert x\vert}}\right) + \\
    & 2^{a-1}{\rm e}^{\gamma\vert x\vert}{\rm B}\left({a+\gamma\over 2}, {a-\gamma\over 2}, {1\over 1+{\rm e}^{2\vert x\vert}}\right).
  \end{split}
\end{equation}
For computational reasons, it can be convenient to express the
incomplete Beta functions above by using their hypergeometric expression
\begin{equation}
{\rm B}(a,b;z)={z^a\over a}\, {}_2F_1(a,1-b,1+a;z).
\end{equation}
We verified the numerical accuracy of the 1-dimensional integration of equations \eqref{eq:besselpot}--\eqref{eq:DExp_rho} by comparison with the potential obtained from equation \eqref{eq:ellpot}.

We conclude by noticing that equation \eqref{eq:besselpot} can be immediately extended to other families of factorised thick discs, with a radial density factor allowing for an explicit Hankel transform. Examples of these thick discs (implemented in JASMINE2), are the Kuzmin disc, the truncated, untruncated and finite Mestel discs, the truncated constant density disc, and the Maclaurin disc (for the relative Hankel transform, see respectively equations 13.148, 2.119, 5.46, 5.55, 5.53, 5.54 in \citealt{ciotti21}; see also \citeauthor{caravita_phd} in preparation).

% Don't change these lines
\bsp	% typesetting comment
\label{lastpage}
\end{document}